# Can fluctuating quantum states acquire the classical behavior on large scale?

Piero Chiarelli

*National Council of Research of Italy, Area of Pisa, 56124 Pisa, Moruzzi 1, Italy*
*Interdepartmental Center "E.Piaggio" University of Pisa*
Phone: +39-050-315-2359
Fax: +39-050-315-2166

Email: pchiare@ifc.cnr.it.

**Abstract:** The quantum hydrodynamic analogy (QHA) equivalent to the Schrödinger equation is derived as a deterministic limit of a more general stochastic version. On large scale, the quantum stochastic hydrodynamic analogy (SQHA) shows dynamics that under some circumstances may acquire the classical evolution. The SQHA puts in evidence that in presence of spatially distributed noise the quantum pseudo-potential restores the quantum behavior on a distance shorter than the correlation length (named here $\lambda_c$) of fluctuations of the quantum wave function modulus. The quantum mechanics is achieved in the deterministic limit when $\lambda_c$ tends to infinity with respect to the scale of the problem. When, the physical length of the problem is of order or larger than $\lambda_c$, the quantum potential (QP) may have a finite range of efficacy maintaining the non-local behavior on a distance $\lambda_L$ (named here "quantum non-locality length") depending both by the noise amplitude and by the inter-particle strength of interaction. In the deterministic limit (quantum mechanics) the model shows that the "quantum non-locality length" $\lambda_L$also becomes infinite. The SQHA unveils that in linear systems fluctuations are not sufficient to break the quantum non-locality showing that $\lambda_L$ is infinite even if $\lambda_c$ is finite.

## 1. Introduction

The emergence of classical behavior from a quantum system is a problem of interest in many branches of physics [1]. The incompatibility between the quantum and classical mechanics comes mainly from the impossibility to manage the non local character of the quantum mechanics. Even if this is a great theoretical problem, from the empirical point of view, the solution seems to be achievable. It has been shown by may authors that fluctuations may destroy quantum coherence and elicit the emergence of the classical behavior [1-6]. By using the alternative approach of the quantum hydrodynamic analogy (QHA) [7] in this paper we investigate how the fluctuations influence the quantum non locality. The goal is to propose a satisfying model that theoretically gives analytical details about the pathway that brings to the quantum decoherence and possibly to the large-scale classical evolution.

The motivation of using the quite unknown QHA can be really appreciated once the overall description is achieved. By now, we can observe that even if the Schrödinger equation is widely known and more manageable than the QHA, it owns some incompatibilities with the large scale local physics: In the Schrödinger approach is not clear how the non-local behavior can be managed to make it compatible with the local character of the classical behavior: non-locality is built-in in the theory and has an infinite range of application. On the other hand, the QHA is practically intractable for any physical problem but it owns a classical-like structure that makes it suitable for the achievement of a comprehensive understanding of quantum and classical phenomena.
The suitability of the classical-like theories in explaining open quantum phenomena is a matter of fact and is confirmed by their success in the description of the dispersive effects in semiconductors [8,9] multiple tunneling [10], mesocopic and quantum Brownian oscillators [11], critical phenomena [12-14], and theoretical regularization procedure of quantum field [15-16].
Since the introduction by Schrödinger of the quantum wave equation, the QHA was presented by Madelung [7] as an alternative equivalent approach to quantum mechanics that gives rise to an interesting logical approach to it.
The interest for the quantum hydrodynamic analogy (QHA) of quantum mechanics had never interrupted since nowadays. It has been studied and extended by many authors as Jánossi [17] resulting useful in the numerical solution of the time-dependent Schrödinger equation [18-20]. More recently it has been used for modeling quantum dissipative phenomena in semiconductors that cannot be described by the semi-classical approximation [8,9].
Moreover, compared to others classical-like approaches (e.g., the stochastic quantization procedure of Nelson [21-23], the mechanics given by Bohm [24-27] and those proposed by Takabayasi [28], Guerra and Ruggiero [29], Parisi and Wu [30] and others [31-32] ) the QHA has the precious property to be exactly equivalent to the Schrödinger equation (giving rise to the same results [18-20]) and it is free from problems such as the unclear relation between the statistical and the quantum fluctuations as in the Nelson theory [21-23] or the undefined variables of the Bohmian mechanics. Concerning the last point, it must be noted that the QHA has not to be confused with the Bohmian mechanics. Even there exist a great similarity between the two theories, as clearly shown by Tsekov [33-34], the Bohmian model seem to be more a mean-field limit of quantum theory than a real punctual model with the defect to possess undefined variables. On the other hand, the QHA has no undefined variables and is perfectly equivalent to the Schrödinger mechanics. The QHA is constituted by two coupled first order differential equation for two real variables: the wave function modulus (WFM) and it phase. By the variable substitution, such a system of equation can be reduced to a single second order differential equation of a complex variable (i.e., the wave function) that is the Schrödinger one.
Among the objectives that could benefit from the present work there are: The clarification of the hierarchy between the classical and quantum mechanics [35-36]; The achievement of a consistent theory of quantum gravity [37-41]; The quantum treatment of chaotic dynamical systems and irreversibility[42-57].
Actually, to describe critical dynamics kinetic Langevin equations are assumed on a phenomenological point of view where it is decided *a priori* what is pertinent to the approximated dynamics. In this context it is really difficult to have a rigorous Langevin description.
The achievement of a theory that on a “small length scale” preserves the standard quantum mechanics while on a large one self-consistently disembogues into the classical one, gives the chance to describe by means of a coarse-grained Langevin equation the connection among irreversibility, chaos and quantum dynamics in a systematic manner by passing from the microscopic scale to the macroscopic one.

## 1.1 Paper outline

In this paper, the standard quantum mechanics (represented in the QHA) is derived as a deterministic limit of a more general stochastic QHA (SQHA). The goal of the work is to show how the quantum mechanics is retrieved in the frame of such a more general theory and under which conditions its non-local character is maintained or modified in the stochastic case. The work investigates how the non-local quantum character (that in the QHA is given by the range of interaction of the quantum pseudo-potential) is restored when the amplitude of the noise converges to zero.
In the case of a small non-zero value of the fluctuations amplitude the work inspects in details: (1) if exists a scale below which the standard quantum mechanics is still achieved (2) what is the range of interaction of the quantum pseudo-potential, (3) how it depends by the fluctuation amplitude and by the inter-particle strength of interaction.

## 2. The SQHA phase space equation of motion

In this section we analyze the QHA in the case of spatially distributed noise. The QHA-equations are based on the fact that the Schrödinger equation, applied to a wave function $\psi_{(q,t)} = A_{(q,t)} \exp[i\, S_{(q,t)} / \hbar]$, is equivalent to the motion of a fluid with particle density $n_{(q,t)} = A^2_{(q,t)}$ and a velocity $\dot{q} = \frac{\nabla_q S_{(q,t)}}{m}$, governed by the equations [ 7]

$$\partial_t n_{(q,t)} + \nabla_q \bullet ( n_{(q,t)} \dot{q} ) = 0 , \tag{1}$$

$$\dot{q} = \nabla_p H , \tag{2}$$

$$\dot{p} = -\nabla_q ( H + V_{qu} ) , \tag{3}$$

with

$$\nabla_p \equiv ( \frac{\partial}{\partial p_1} ,...., \frac{\partial}{\partial p_{3n}} ) \tag{4}$$

$$\nabla_q \equiv ( \frac{\partial}{\partial q_1} ,...., \frac{\partial}{\partial q_{3n}} ) \tag{5}$$

where

$$H = \frac{p \bullet p}{2m} + V_{(q)} \tag{6}$$

is the Hamiltonian of the system of *n* structureless particles of mass *m* and $V^{qu}$ is the quantum pseudo-potential that reads

$$V_{qu} = -( \frac{\hbar^2}{2m} ) n^{-1/2} \nabla_q \bullet \nabla_q n^{1/2} . \tag{7}$$

For the purpose of this paper, it is useful to observe that equations (1-3) can be derived by the following phase-space equations

$$\partial_t {}^\circ_{(q,p,t)} + \nabla \bullet ( {}^\circ_{(q,p,t)} ( \dot{x}_H + \dot{x}_{qu} ) ) = 0 \tag{8}$$

where

$$n_{(q,t)} = \iiint {}^\circ_{(q,p,t)} d^{3n} p . \tag{9}$$

$$\dot{x}_H = ( \nabla_p H , -\nabla_q H ) \tag{10}$$

$$\dot{x}_{qu} = ( 0, - \nabla_q V_{qu} ) \tag{11}$$

once equation (8) is integrated over the momentum *p* with the sufficiently general condition that

$$\lim_{|p| \to \infty} {}^\circ_{(q,p,t)} = 0 \tag{12}$$

and the phase space quantum field ° has the form

$$ {}^{\circ}{}_{(q,p,t)} = \mathrm{n}_{(q,t)} ¢( p - \nabla_q S ) \quad , \tag{13} $$

where

$$ S = \int_{t_0}^{t} dt( \frac{p \bullet p}{2m} - V_{(q)} - V_{qu} ) . \tag{14} $$

Due to the fact that the ensemble of solutions of equations (8-11) is wider than that one of the QHA-equations (1-3), the accessory condition

$$ \dot{q} = \frac{\nabla_q S}{m} \tag{15} $$

(namely the wave-particle equivalence) warranted by the δ-function in (6) must be applied to (8)- Generally speaking, for a solutions of the problem (8) we have

$$ \dot{q}_{(q,t)} = \frac{\iiint \dot{q}_{(q,p,t)} \, {}^{\circ}{}_{(q,p,t)} d^{3n} p}{\mathrm{n}_{(q,t)}} \neq \frac{\nabla_q S}{m} \tag{16} $$

This is an important point since the satisfaction of condition (15) is necessary to pass back from the QHA equations to the Schrödinger one [7,55].
For the more general case of spatially distributed stochastic noise, the stochastic-PDE (SPDE), whose zero noise limit can lead to the deterministic PDE (4), reads

$$ \partial_t {}^{\circ}{}_{(q,p,t)} = -\nabla \bullet ( {}^{\circ}{}_{(q,p,t)} ( \dot{x}_H + \dot{x}_{qu} )) + ¦_{(q,t,\Theta)} ¢( p - \nabla_q S ) , \tag{17} $$

where Θ is a measure of the noise amplitude and where the accessory condition (see Appendix A)

$$ {}^{\circ}{}_{(q,p,t)} = \mathrm{n}_{(q,t)} ¢( p - \nabla_q S ) , \tag{18} $$

is held in order to warrants the wave particle equivalence in the deterministic limit.

## 2.1 Perturbing expansion around the deterministic quantum mechanics

Since in the limit of zero noise the ensemble of solution of the SPDE (17) is wider than that one of the deterministic equation (8) (see for instance Ref. [56]), it needs to enucleate the conditions that warrant the establishing of the quantum mechanics (i.e., the PDE (1)) for Θ that goes to zero.
To this end, we investigate (17) in the limit of small noise amplitude Θ for the sufficiently general case to be of great interest of a Gaussian random noise.
In order to perturbingly investigate (17) near the deterministic limit, we re-write it as

$$ \partial_t {}^{\circ}{}_{(q,p,t)} = -\nabla \bullet ( {}^{\circ}{}_{(q,p,t)} ( \dot{x}_H + \dot{x}_{qu( {}^{\circ}{}_0 )} + F^* )) + ¦_{(q,t,\Theta)} ¢( p - \nabla_q S ) , \tag{19} $$

where $°_0$ is the solution of the PDE (8) and where $F^*$(x,t), containing the QP fluctuations due to the SPDE field ρ (that can be very large even in presence of a vanishing noise) reads

$$ F^* = \nabla I^* , \tag{20} $$

$$I^* = -(\frac{\hbar^2}{2m})\{ n^{-1/2}\nabla_q \bullet \nabla_q n^{1/2} - n_0{}^{-1/2}\nabla_q \bullet \nabla_q n_0{}^{1/2} \} \\ = V_{qu(n)} - V_{qu(n_0)} \quad . \tag{21}$$

and where

$$n_{0(q,t)} = \iiint {}^{\circ}{}_{0(q,p,t)} d^{3n} p \, . \tag{22}$$

$$\partial_t {}^{\circ}{}_{0(q,p,t)} + \nabla \bullet ( {}^{\circ}{}_{0(q,p,t)} ( \dot{x}_H + \dot{x}_{qu} )) = 0 \tag{23}$$

Equation (21) without the term $\nabla \bullet ( {}^{\circ}{}_{(q,p,t)} F^*)$ is a flow equation with spatially distributed noise that has been already extensively studied in the form a Fokker-plank equation [57] that converges to the deterministic limit for Θ going to zero.
Actually, the derivative structure of the term $\nabla \bullet ( {}^{\circ}{}_{(q,p,t)} F^*)$, in principle can lead to a finite contribution even if the noise amplitude Θ is vanishing. This is quite evident since a small abrupt variation of the quantum field ${}^{\circ}{}_{(q,p,t)}$ can give a large output on its derivative in the quantum potential expression.
In principle, nothing makes the solution expressing the quantum mechanics in (19) privileged except the physical constraints that introduced into the abstract plane of the equations will explain why all alternative solutions cannot happen but those of quantum mechanics.
Following this logical pathway, we observe that in order the fluctuating states of the SPDE (21) can realizes themselves, it is necessary that their energy is finite, that is: The energy gap between the quantum deterministic states of (1) and the corresponding fluctuating one has to be finite. Given that the energy of the QP is a real energy for the system [58], if we impose that the energy of the fluctuating state is finite, we have also to impose that the energy increase introduced by the fluctuations of the QP is finite, that is

$$\lim_{\Theta \to 0} < (V_{qu} - V_{qu(n_0)})^2 >= 0 \tag{25}$$

(more precisely, we impose that the root mean square of QP energy fluctuations in (20) are finite).
As far as it concerns this point, in presence of spatially distributed noise, the derivative structure of the QP (3) immediately shows to play an important role since the fluctuations of the quantum field °, on shorter and shorter distance, will produce higher and higher QP values. Therefore, since white quantum field fluctuations would lead to an infinite value of the QP energy fluctuations, they are not possible for the problem (19). Moreover, the fact that spatially non-white quantum field fluctuations are the consequence of the action of the QP, means that it acts to suppress them on shorter and shorter distance. This QP cut-off of high spatial quantum field fluctuations frequencies clearly means that exists a distance $\lambda_c > 0$ (we name it here “quantum coherence length”), below which the noise is damped and the standard quantum mechanics (deterministic limit) are approached.
For sake of completeness, we have to observe that if we want to build up a perturbing expansion around the deterministic field solution (23), making Θ a true perturbing expansion parameter in the sense that

$$\lim_{\Theta \to 0} < ( {}^{\circ}{}_{(q,p,t)} - {}^{\circ}{}_{0(q,p,t)} )^2 >= 0 \, , \tag{26}$$

actually, we have to warrant that the overall term $\nabla \bullet ( {}^{\circ}{}_{(q,p,t)} F^*)$ converge to zero in the sense

$$\lim_{\Theta \to 0} < (\nabla_q \circ F^*)^2 >= 0 \qquad (27)$$

As shown in the appendix B, the energy requirement given by (25) is a sufficient condition to warrant (26).

On the condition (to be defined) that the root mean square of the QP energy fluctuations is finite (for any $\Theta$ finite), so that equation (19) converges in the sense of (27) to the deterministic limit for $\Theta$ going to zero, it is possible to approximate the field solution (whose integral over momenta gives the WFM) as the sum in the following

$$\circ \cong \circ_0 + \Delta\circ_1 + \Delta\circ_2 + .......... + \Delta\circ_n + ....... \qquad (28)$$

so that the m-th order solution reads

$$\circ_m \cong \circ_0 + \Delta\circ_1 + \Delta\circ_2 + .......... + \Delta\circ_m . \qquad (29)$$

Moreover, given the general initial conditions

$$\circ_{m(q,p,t=0)} = \circ_0 \qquad \forall \mathrm{m} > 0 \qquad (30)$$

it will hold

$$\lim_{t \to 0} \Delta\circ_{m(q,p,t)} = 0 \qquad \forall\, m \qquad (31)$$

and hence, under the condition of Gaussian field fluctuations (to be check at the end) in the vanishing noise amplitude $\Theta$ , for a sufficiently short interval of time $\Delta$t after the initial instant, the solution $\circ_m$ will remain close to $\circ_{m-1}$ in the sense that

$$\lim_{t \to 0} < (\circ_{m(q,p,t)} - \circ_{m-1(q,p,t)})^2 >= 0 . \qquad (32)$$

In force of that, in such an interval of time $\Delta$t, the equation of motion (19) can be approximated by the following system of coupled equations

$$\partial_t \circ_{0(q,p,t)} \cong -\nabla(\circ_{0(q,p,t)}(\dot{x}_H + \dot{x}_{qu(\circ_0)})), \qquad (33)$$

$$\partial_t \circ_{1(q,p,t)} \cong -\nabla(\circ_{1(q,p,t)}(\dot{x}_H + \dot{x}_{qu(\circ_0)})) + N_1 \qquad (34)$$

$$\partial_t \circ_{m(q,p,t)} \cong -\nabla(\circ_{m(q,p,t)}(\dot{x}_H + \dot{x}_{qu(\circ_0)})) + N_m \qquad (35)$$

and so on; where

$$N_m = -\nabla \bullet (\circ_{m(q,p,t)} F^*_m) + ¦_{(q,t,\Theta)} ¢ (p - \nabla_q S) \qquad (36)$$

$$F^*_m \cong \nabla(V_{qu(n_{m-1})} - V_{qu(n_0)}) \qquad (37)$$

$$n_{m(q,t)} = \iiint \circ_{m(q,p,t)} d^{3n} p \qquad (38)$$

$$S_m = \int_{t_0}^{t} dt\left(\frac{p \bullet p}{2m} - V_{(q)} - V_{qu(n_m)}\right) \qquad (39)$$

and where the approximation (32) has been introduced into (37).
Furthermore, given that from (37) it follows that $F^*_1 = 0$, equation (34) at first order of approximation reads

$$\partial_t \rho_{1(q,p,t)} \cong -\nabla(\rho_{1(q,p,t)}(\dot{x}_H + \dot{x}_{qu(\rho_0)})) + \eta_{(q,t,\Theta)}\delta(p - \nabla_q S) \qquad (39)$$

We note that the approximation in equation (39) $F^*_1 \cong 0$, steaming by (37), is correct providing that in the small noise amplitude the condition (25) and hence (26) are satisfied.
After the interval of time $\Delta$t is passed by, the final condition $\rho_{1(q,p,t+\Delta t)}$ is assumed as the initial condition for the subsequent interval of time $\Delta$t. The repetition of such a procedure will allow to find the solution at an arbitrary instant of time into the future for $\Theta$ sufficiently close to zero.
If we consider the sufficiently general case to be of practical interest that $N_1$ is Gaussian, in order to determine it, we need to define its mean and correlation function. If we assume that: (1) the noise has null mean; (2) The noise has null correlation time; (3) The space is isotropic; (4) the noises on different co-ordinates are independent; $\eta_{(q,t,\Theta)}$ generally reads

$$<\eta_{(q_\alpha,t)}, \eta_{(q_\beta+\lambda,t+\tau)}> = \delta_{\alpha\beta}\, g(0) G(\lambda)\delta(\tau) \qquad (40)$$

At lowest order the SQHA problem reduces to find the noise shape G($\lambda$) that warrants conditions (25,26). As shown in Appendix B, this is obtained by requiring that the QP energy does not diverge when the correlation distance of the WFM fluctuations tends to zero. This can be implemented in the small noise approximation, by using the WFM fluctuations at the zero order of approximation (see (C.31) in appendix C).

### 2.1.2. Fluctuations amplitude of the quantum potential energy

Here we derive the conditions, on the correlation length of the zero order Gaussian WFM fluctuations, under which the QP root mean square energy fluctuations become vanishing and the convergence to the deterministic PDF (1) is warranted for $\Theta$ vanishing.
Once the (lowest order) WFM fluctuations are defined by (57-40) (see (C.31) in appendix C) we can correspondingly evaluate the QP fluctuations.
Given the zero order Gaussian WFM fluctuations (C.31), we can find the form of $G(\lambda)$ by imposing that the root mean square of QP fluctuations do not diverge as $\Theta$ goes to zero. This is practically implemented by operating in the discrete approach and then by passing to the continuous limit (see appendix C and D).
Since in the discrete approach the correlation length of fluctuations cannot be smaller than the discrete cell dimension $\lambda$, the non-diverging QP condition is practically obtained by imposing that the root mean square of QP fluctuation, in the discrete form, remains limited when the cell dimension $\lambda$ goes to zero.
By using the QP expression that after simple manipulation reads

$$V_{qu} = -(\frac{\hbar^2}{2m})[n^{-1}\nabla_q \bullet \nabla_q n - n^{-2}\nabla_q n \bullet \nabla_q n], \qquad (41)$$

and by writing the spatial quantum field derivatives as the limit of the corresponding discrete quantity as

$$\nabla_q n \bullet \nabla_q n = \lim_{\lambda \to 0} \sum_{\alpha} \lambda^{-2}[n_{(q_{\alpha(i+1)})} - n_{(q_{\alpha(i)})}]^2 \qquad (42)$$

and

$$\nabla_q \bullet \nabla_q n = \lim_{\lambda \to 0} \sum_{\sim} \lambda^{-2} [\, n_{(q_{\sim(i+2)})} - 2n_{(q_{\sim(i+1)})} + n_{(q_{\sim(i)})} \,] \,, \qquad (43)$$

we are able to calculate the variance of the QP fluctuations (41) by means of the following equalities [see Appendix D]

$$\lim_{\lambda \to 0} < \nabla_q n \bullet \nabla_q n, \nabla_q n \bullet \nabla_q n > \; = \; \lim_{\lambda \to 0} \lambda^{-4} \{ \sum_{\sim} 4\Delta t^2 \, g(0)^2 [1 - G(\lambda)] \}^2$$

$$+ \lim_{\lambda \to 0} \lambda^{-2} \sum_{\sim} 4A\Delta t \, g(0)[1 - G_{\sim}(\lambda)] \,,$$

(44)

$$\lim_{\lambda \to 0} < \nabla_q \bullet \nabla_q n, \nabla_q \bullet \nabla_q n >= \lim_{\lambda \to 0} \lambda^{-4} \sum_{\sim} 2\Delta t \, g(0)[3 + G(2\lambda) - 4G(\lambda)]$$

(45)

and

$$\lim_{\lambda \to 0} < \nabla_q \bullet \nabla_q n, \nabla_q n \bullet \nabla_q n >= 0 \,, \qquad (46)$$

where

$$g(0) = < \text{¦}_{(q_\sim)}, \text{¦}_{(q_\sim)} > \,, \qquad (47)$$

$$G(\lambda) = \frac{< \text{¦}_{(q_\sim)}, \text{¦}_{(q_\sim + \lambda)} >}{g(0)} \qquad (48)$$

$$G(2\lambda) = \frac{< \text{¦}_{(q_\sim)}, \text{¦}_{(q_\sim + 2\lambda)} >}{g(0)} \,. \qquad (49)$$

$$A = < \frac{\partial n^0}{\partial q_\sim} > < \frac{\partial n^0}{\partial q} >$$

$$\Delta t = t - t_0$$

and where in (69) has been used the identity $\lim_{\lambda \to 0} < \text{¦}_{(q_\sim)} >= 0$.

If we require that the root mean square of the QP energy fluctuations (see Appendix D) satisfies the condition

$$\lim_{\lambda \to 0} < V_{qu}, V_{qu} >$$

$$= \lim_{\lambda \to 0} \left(\frac{\hbar^2}{2m}\right)^2 [\, d_1^{-2} < \nabla_q \bullet \nabla_q n, \nabla_q \bullet \nabla_q n > - d_2^{-4} < \nabla_q n \bullet \nabla_q n, \nabla_q n \bullet \nabla_q n > ] < \infty$$

(50)

when $\lambda \to 0$ (where $d_1$ and $d_2$ are weighted mean particle densities given in Appendix D), it must necessarily follow that

$$\lim_{\lambda \to 0} \sum_{\alpha} \lambda^{-2} [1 - G_{\alpha}(\lambda)] < \infty \tag{51}$$

$$\lim_{\lambda \to 0} \sum_{\alpha} \lambda^{-4} [1 - G_{(\lambda)}]^{2} < \infty \tag{52}$$

$$\lim_{\lambda \to 0} \sum_{\alpha} \lambda^{-4} [3 + G_{(2\lambda)} - 4G_{(\lambda)}] < \infty \tag{53}$$

Developing $G_{\alpha}(\lambda)$ for small $\Theta$ in series expansion as a function of $\lambda/\lambda_c$, where $\lambda_c$ is defined further on, we obtain

$$\lim_{\lambda \to 0} G_{(\lambda)} \cong a_0 + a_1 \frac{\lambda}{\lambda_c} + a_2 (\frac{\lambda}{\lambda_c})^2 + a_3 (\frac{\lambda}{\lambda_c})^3 + a_4 (\frac{\lambda}{\lambda_c})^4 + \sum_{j=1}^{\infty} a_j (\frac{\lambda}{\lambda_c})^j \,, \tag{54}$$

from where it follows that (72-3 51-53) are verified if $a_0 = 1$, $a_1 = 0$, and $a_3 = 0$, while no condition applies to the coefficients $a_2$ and $a_j$ with $j \geq 4$ that are unable to produce the divergence of (51-53) and remain undefined. Therefore, $G(\lambda)$ reads

$$\lim_{\lambda \to 0} G_{(\lambda)} \cong 1 + a_2 (\frac{\lambda}{\lambda_c})^2 + a_4 (\frac{\lambda}{\lambda_c})^4 + \sum_{j=1}^{\infty} a_j (\frac{\lambda}{\lambda_c})^j \,. \tag{55}$$

where without a leaking of generality we can put $a_2 = \pm 1$ by a re-definition of the spatial cell side $\lambda$ such as $\lambda' = a_2^{1/2}\lambda$. Introducing (55) into (51-53) for a check, we obtain

$$\lim_{\lambda \to 0} \sum_{\alpha} \lambda^{-2} [1 - G_{\alpha}(\lambda)] = -\frac{a_2}{\lambda_c^{\,2}} \tag{56}$$

$$\lim_{\lambda \to 0} \sum_{\alpha} \lambda^{-4} [1 - G_{(\lambda)}]^{2} = \frac{a_2^{\,2}}{\lambda_c^{\,4}} \tag{57}$$

$$\lim_{\lambda \to 0} \sum_{\alpha} \lambda^{-4} [3 + G_{(2\lambda)} - 4G_{(\lambda)}] = \frac{16 a_4}{\lambda_c^{\,4}} \tag{58}$$

.

## 2.2. Extension of the SQHA model to the macroscopic scale ( $\lambda >> \lambda_c$ )

As shown further on in this paper, since the quantum coherence length $\lambda_c$ results by the geometrical mean of the Compton length $l_C$ and the stochastic length $\frac{\hbar c}{k\Theta}$, the SQHA model is anyway linked to the Compton length as a reference scale.

Therefore, for a macroscopic system whose dimensions are huge compared to the quantum coherence length $\lambda_c$ (for instance, for $\Theta$ as small as 1°K, it results $\lambda_c = (l_C \hbar c / k\Theta)^{½} = (2.2 \times 10^{-13})^{½} = 4.7 \times 10^{-6}$ cm for a particle of proton mass) the continuum macroscopic limit can be achieved for λ that numerically goes to zero but that is still very large compared to $\lambda_c$. In this case it is not enough to know the series expansion (55) of the correlation function .
Given the physical values of the Compton's length $l_C$, very small noise amplitude Θ (of order of one or tens of degree Kelvin) still leads to a very small value of $\lambda_c$ compared with the standard length units of the macroscopic physics. Therefore in this case, the Gaussian small noise approximation can hold without solution of continuity from the micro-scale to the macro-scale approaches.
In order to obtain a model holding also for a large-scale approach, hence, we utilize the correlation function of the Gaussian fluctuations in the form of an exponential law in agreement with (73 55).
To this end, we investigate in detail the model with $a_2 = -1$ (that warrants the ergodicity) in the particular case where the shape of the correlation function reads

$$G_{(a)} = exp[-(\frac{a}{a_c})^2] \ . \qquad (59)$$

that both satisfies (55) and leads to a large-scale δ-correlated spatial fluctuations.
The model with Gaussian quantum field fluctuations that owns (59) as correlation function does not exclude others macroscopic Gaussian noises that are present at the large-scale level [59].
Given that the Gaussian processes $¦_{(q)}$ with the correlation function

$$< ¦_{(q_{\sim},t)}, ¦_{(q+a,t+²)} >= g(0)exp[-(\frac{a}{a_c})^2]¢(²)¢_{\sim} \ , \qquad (60)$$

where (see appendix E ) $g(0)$ and $a_c$ respectively read

$$g(0) =< ¦_{(q_{\sim})}, ¦_{(q_{\sim})} > \propto \frac{k\Theta}{a_c^{\ 2}} , \qquad (61)$$

$$a_c = (\frac{®}{2})^{3/2} \frac{\hbar}{(2mk\Theta)^{1/2}} = (\frac{®}{2})^{3/2} (\frac{l_C \hbar c}{2k\Theta})^{1/2} \qquad (62)$$

where $l_C = \frac{\hbar}{mc}$ is the Compton's length, the large scale correlation function reads

$$\lim_{a/a_c \to \infty} < ¦_{(q_{\sim},t)}, ¦_{(q+a,t+²)} > \propto \lim_{a/a_c \to \infty} ¢_{\sim} \frac{k\Theta}{a_c^{\ 2}} exp[-(\frac{a}{a_c})^2]¢(²)¢_{\sim}$$
$$\propto ¢_{\sim} \frac{2k\Theta}{a_c} ¢(a)¢(²) \qquad (63)$$

In the Appendix E, the quantum coherence length $a_c$ is calculated by imposing that the root mean square of the system energy fluctuation at equilibrium calculated by the SQHA are the same to that one obtained by statistical mechanics. The expression of $a_c$ is obtained with the convention that the vacuum fluctuations amplitude Θ (in the small value limit) is measured in a scale by which it equals the temperature of an ideal gas at equilibrium (placed in such a space region).
By using (62), finally, (62) reads

$$g(0) =< ¦_{(q_{\sim})}, ¦_{(q_{\sim})} >= « \frac{8m(k\Theta)^2}{®^3 \hbar^2} \qquad (64)$$

where the “form“ factor $\underline{\text{«}}$ for an ideal gas confined in a vessel of side $\Delta L$ reads (see Appendix E)

$$\underline{\text{«}} = \frac{\text{«}}{\Delta L^6} \qquad (65)$$

where

$$\text{«} = \frac{\Delta L^2}{\hbar} \qquad (66)$$

has the dimension of a mobility constant. Thence, the motion equation in the small noise approximation reads

$$\partial_t \text{n}_{(q,t)} = -\nabla_q \bullet (\text{n}_{(q,t)} \dot{q}_{(\text{n}_0)}) + \text{¦}_{(q_{\sim},t,\Theta)} \qquad (67)$$

$$< \text{¦}_{(q_{\sim},t)}, \text{¦}_{(q_{\sim}+\text{a},t+\text{²})} >= \underline{\text{«}} \frac{8m(k\Theta)^2}{\text{®}^3 \hbar^2} exp[-(\frac{\text{a}}{\text{a}_c})^2]\text{¢}(\text{²})\text{¢}_{\sim} \qquad (68)$$

$$\text{a}_c = (\frac{\text{®}}{2})^{3/2} \frac{\hbar}{(2mk\Theta)^{1/2}} \qquad (69)$$

$$\dot{q}_{(\text{n}_0)} = \frac{p_0}{m}, \qquad (70)$$

$$\dot{p}_0 = -\nabla_q (V_{(q)} + V_{qu(\text{n}_0)}), \qquad (71)$$

$$\partial_t \text{n}_{0(q,t)} = -\nabla_q \bullet (\text{n}_{0(q,t)} \dot{q}_{(\text{n}_0)}) \qquad (72)$$

$$S = \int_{t_0}^{t} dt(\frac{p \bullet p}{2m} - V_{(q)} - V_{qu(\text{n})}) = \int_{t_0}^{t} dt(\frac{p \bullet p}{2m} - V_{(q)} - V_{qu(\text{n}_0)} - I^*) \qquad (73)$$

$$m\dot{q} = p = \nabla_q S = \nabla_q \{ \int_{t_0}^{t} dt(\frac{p \bullet p}{2m} - V_{(q)} - V_{qu(\text{n}_0)} - I^*)\} = p_0 + \Delta p_{st} \qquad (74)$$

where

$$\Delta p_{st} = -\nabla_q \{ \int_{t_0}^{t} I^* dt \}, \qquad (75)$$

#### 2.2.1. Large-scale quantum force

In addition to the large-scale noise limit, to obtain the macro-scale form of equations (67-75) we need to investigate the large-scale limit of the quantum force $\dot{p}_{qu} = -\nabla_q V_{qu}$ in (71).

The behavior of the WFM determines the QP through the term $n^{1/2}$ of (3). For sake of simplicity we discuss here the one-dimensional case of localized state that at large distance $n^{1/2}$ goes like

$$m\text{-}s \lim_{|q|\to\infty} n^{1/2} \propto exp[-P^{h}{}_{(q)}] \qquad (76)$$

where $P^{h}{}_{(q)}$ is a polynomial of degree equal to $h$, $z_q = \gamma^{-1}q$ is the macroscopic variable (where $\gamma = \Delta\Omega q/\lambda_L$ and $\Delta\Omega q$ is the macro-scale resolution cell size) and $\lambda_L$ is the range (to be defined) of interaction of the QP.
The QP (3) at large scale reads

$$\lim_{¥\to\infty} V_{qu} = \lim_{¥\to\infty} -h^2 ¥^{2(h-1)} z_q^{2(h-1)} + h(h-1)¥^{h-2} z_q^{h-2} \qquad (77)$$

Changing the exponent variable by $\phi = 3 - 2h$, we obtain

$$\lim_{¥\to\infty} V_{qu} = \lim_{¥\to\infty} -h^2 ¥^{-¤} z_q^{1-¤} + h(h-1)¥^{-(1.5+¤)} z_q^{-(3+¤)/2} \qquad (78)$$

and therefore the quantum force $-\nabla_q V_{qu}$ at large scale reads

$$\lim_{¥\to\infty} -\nabla_q V_{qu} = \lim_{¥\to\infty} h^2 2(h-1)(¥z_q)^{-¤} + h(h-1)(h-2)(¥z_q)^{-½(3+2¤)} z_q^{½¤} \qquad (79)$$

that for $\phi > 0$ (i.e., $h < 3/2$) $\forall\, z_q \neq 0$ gives

$$\lim_{q\to\infty(z_q\ finite)} -\nabla_q V_{qu} = \lim_{q\to\infty(z_q\ finite)} 2h^2(h-1)q^{-¤} + h(h-1)(h-2)q^{-½(3+2¤)} z_q^{½¤}$$
$$\approx 2h^2(h-1)q^{-¤} = 0 \qquad (80)$$

Moreover, since the following integral

$$\int_0^{\infty} |q^{-1}\nabla_q V_{qu}|\,dq \propto \int_0^{\infty} |\frac{1}{q^{1+¤}}|\,dq < \infty \qquad (81)$$

converges for $\phi > 0$ (i.e., $h < 3/2$), the requisite (81) tells us whether or not the QP force is negligible on large scale as given by (80).
It is worth mentioning that condition (81) is not satisfied for linear system whose eigenstates have $h = 2$ so that they cannot admit the classical limit.
It is also worth noting that condition (80), obtained for a WFM owing the form (76), holds also in the case of oscillating wave functions that at large distances are of type

$$\lim_{|q|\to\infty} |\cdot| = \lim_{|q|\to\infty} n^{1/2} = M_{(q)}\, exp[-P^{h}{}_{(q)}] \qquad (82)$$

with

$$\lim_{|q|\to\infty} M_{(q)} \cong q^m \sum_n a_n\, exp[iA_n{}^{p}{}_{(q)}] \qquad (83)$$

where $A_n{}^{p}{}_{(q)}$ are polynomials of degree equal to $p$. In this case, in addition to the requisite

$$0 \le h < \frac{3}{2} \tag{84}$$

the conditions $m \in \Re$ and $p \le 1$ are required (see appendix F).
Moreover, given that for regular continuous and derivable Hamiltonian forces that are attractive at large distance and whose zero potential can be put to infinity (already sufficiently general to be of great practical interest such as the L-J type potentials) it holds

$$\lim_{|q| \to \infty} A_n{}^{p}{}_{(q)(q)} \propto q \tag{85}$$

so that in this case $p = 1$ and hence (81) is still a valid condition in order to warrant the existence of a finite $a_L$.
Finally it must be observed that even the convergence (76) is satisfied, there can exist short length quantum wave modulus oscillations that can leads to very high QP values (that in principle cannot be disregarded). Respect to this we observe that: (1) since any curvature of the quantum wave modulus generates a quantum potential force that opposes itself to it like an “elastic type” response, such short-length oscillations were smoothed out in a short interval of time. Since in a large –scale description we are not interested in microscopic details as well as in a very short time behavior, they can be disregarded. (2) the integral condition (81) is not influenced by this micro-scale oscillations since it mediates over such oscillations so that it is able to discriminate if the global behavior of the QP is relevant at large distances (bigger than $a_c$).

### 2.2.2. Quantum non-locality length $a_L$

By considering (81) as a measure of the quantum potential force at large distance, the quantum potential range of interaction can be obtained as the mean weighted distance of the integrand of (81) that for the unidimensional case reads

$$a_L = 2 \frac{\int_0^\infty | q^{-1} \frac{dV_{qu}}{dq} | dq}{a_c{}^{-1} | \frac{dV_{qu}}{dq} |_{(q = a_c)}} . \tag{86}$$

where the origin (0,0) is the mean position of the particle. When $a_L \to \infty$, with $\lambda_c$ finite, so that

$$\int_0^\infty | q^{-1} \frac{dV_{qu}}{dq} | dq \to \infty \tag{87}$$

(e.g., it happens for Gaussian states of linear systems), the quantum potential is not vanishing at infinite and the system evolution is affected by the quantum non-local forces even on large scale dynamics.
Given the 3n-dimension generalization of (81) leading to the scalar parameter $F_{qu(r, g_1, ...., \ddot{}_{3n-1})}$ measuring the non-local strength of the quantum force

$$F_{qu(r,g_1,....,\ddot{}_{3n-1})} = (\frac{3n}{2})^{-1} \int_0^{r_1} ... \int_0^{r_{3n}} |(q \bullet q)^{-1/2} \vec{n} \bullet \nabla_q V_{qu} | \bullet d|q|$$
$$= (\frac{3n}{2})^{-1} \int_0^{r_1} ... \int_0^{r_{3n}} |(q \bullet q)^{-3/2} \vec{q} \bullet \nabla_q V_{qu} | \vec{q} \bullet d\vec{q} \qquad (88)$$

where $\vec{n} = \frac{\vec{q}}{|q|}$, the quantum non-locality length $a_L$ can be defined as

$$a_L = a_c \, max\{ \frac{\lim_{r \to \infty} F_{qu(r,g_1,...,\ddot{}_{3n-1})}}{(\nabla_q V_{qu} \bullet \nabla_q V_{qu})^{1/2}{}_{(r = a_c, g_1,...,\ddot{}_{3n-1})}} \} \qquad (89)$$

The expression (89) for a system of a large number of particles is quite complex, nevertheless for the interaction of a couple of particles (e.g., mono-dimensional case, real gas or a chain of neighbors interacting atoms) the expression (86) is quite manageable [60].

### 2.2.3. Limiting dynamics

Since the SPDE (67) depends by two lengths $\lambda_c$ and $a_L$, various limiting cases are possible. In order to correctly name and identify them, in the following we use adjectives according to the rule: (1) $\Delta\Omega q << \lambda_c$ "microscopic". (2) $\Delta\Omega q >> \lambda_c$ "macroscopic". (3) $\Delta\Omega_L << a_L$ "non-local". (4) $a_L << \Delta\Omega q$ "local". (5) $k\Theta \neq 0$ "stochastic". (6) $k\Theta = 0$, "deterministic". (7) $\lambda = \Delta\Omega q$ "continuum limit", where $\Delta\Omega q$ is the resolution length in the scale of the problem and $\Delta\Omega_L$ (with $\Delta\Omega_L >> \Delta\Omega q$ ) is the physical length of the system.

### 2.2.4. Deterministic dynamics

1) *Non-local deterministic dynamics* (i.e., the standard quantum mechanics). The present case is identified by the following values of the parameters $k\Theta = 0$ (i.e., $\lambda_c = \infty$, $a_L = \infty$).
Given by (81) that

$$\lim_{\Theta \to 0} < \dot{\imath}_{(q_\sim, t)}, \dot{\imath}_{(q_\sim + a, t+2)} > = \lim_{\Theta \to 0} ¢_\sim \ll \frac{k\Theta}{a_c^2} exp[-(\frac{a}{a_c})^2] ¢(2) ¢_\sim = 0$$

it follows that the SPDE (67-74) assumes the deterministic form

$$\partial_t n_{0(q,t)} = -\nabla_q \bullet (n_{0(q,t)} \dot{q}) \qquad (90)$$

$$S = \int_{t_0}^{t} dt(\frac{p \bullet p}{2m} - V_{(q)} - V_{qu(n_0)}) = S_{qu} \qquad (91)$$

$$p = \nabla_q S = \nabla_q \{ \int_{t_0}^{t} dt(\frac{p \bullet p}{2m} - V_{(q)} - V_{qu(n_0)}) \} = \nabla_q S_{qu} = p_0 \qquad (92)$$

$$\dot{q} = \frac{p}{m}, \qquad (93)$$

$$\dot{p} = -\nabla_q (V_{(q)} + V_{qu(n_0)}), \tag{94}$$

2) *Local (classic) deterministic dynamics.* Even if the local deterministic dynamics requires $a_L$ = 0 being $\lambda_c$=∞, it is interesting for confined particles (i.e., $\lim_{|q|\to\infty} \rho = 0 \cup \lim_{|q|\to\infty} \partial\rho/\partial x_j = 0$) to discuss this case.

Given that the deterministic dynamics requires kΘ = 0 ($\lambda_c$ = ∞) and that the local character requires $a_L$ = 0, by (93) it is straightforward to demonstrate that for a system of confined particles this can happen if and only if $\hbar = 0$.

The result is immediately achieved by observing that in order to have $a_L$ = 0 (for $\lambda_c$ tending to ∞), it must indeed result $\int_0^\infty |q^{-1} \frac{dV_{qu}}{dq}| dq = 0$ and hence $|\frac{dV_{qu}}{dq}|$ rigorously null over all the space.

For confined particles for which it holds $\lim_{|q|\to\infty} \rho = 0$ and $\lim_{|q|\to\infty} \partial\rho/\partial x_j = 0$, a rigorously constant QP (i.e., $|\partial V^{qu}(q)/\partial q| = 0$) can be obtained from (3) just for the meaningless case ρ = 0 over all the space if $\hbar \neq 0$.

(therefore, it follows that since $\hbar$ is a not null physical constant, in the SQHA model the classical behavior cannot happen in the deterministic manner for localized particles but only in the stochastic mode.

**2.2.5. Stochastic dynamics**

1) *Macroscopic non-local stochastic dynamics*. The present case is identified by the following values of the parameters kΘ ≠ 0 with $\lambda_c$ << ΔΩq <<$\Delta\Omega_L$ << $a_L$ or even $a_L$ = ∞ (e.g., Gaussian states of a linear system). In this case the system of equations (67-74) read

$$\partial_t n_{(q,t)} = -\nabla_q \bullet (n_{(q,t)} \dot{q}_{(n_0)}) + ¦_{(q_{\sim},t,\Theta)} \tag{95}$$

$$< ¦_{(q_{\sim},t)}, ¦_{(q_{\sim}+a,t+^2)} > = \underline{«} ¢_{\sim} \frac{2k\Theta}{a_c} ¢(a) ¢(^2) \tag{96}$$

$$\dot{q}_{(n_0)} = \frac{p_0}{m}, \tag{97}$$

$$\dot{p}_0 = -\nabla_q (V_{(q)} + V_{qu(n_0)}), \tag{98}$$

$$\partial_t n_{0(q,t)} = -\nabla_q \bullet (n_{0(q,t)} \dot{q}_{(n_0)}) \tag{99}$$

$$S = \int_{t_0}^{t} dt(\frac{p \bullet p}{2m} - V_{(q)} - V_{qu(n)}) = \int_{t_0}^{t} dt(\frac{p \bullet p}{2m} - V_{(q)} - V_{qu(n_0)} - I^*) = S_{qu} + S_{st} \tag{100}$$

$$S_{st} = -\int_{t_0}^{t} I^* \, dt \qquad (101)$$

$$m\dot{q} = p = \nabla_q S = \nabla_q \{ \int_{t_0}^{t} dt( \frac{p \bullet p}{2m} - V_{(q)} - V_{qu(\mathrm{n}_0)} - I^*) \} = p_{cl} + \Delta p_{qu} + \Delta p_{st}$$

(102)

where

$$p_{cl} = -\nabla_q \int_{t_0}^{t} V_{(q)} dt \qquad (103)$$

$$\Delta p_{qu} = -\nabla_q \int_{t_0}^{t} V_{qu(\mathrm{n}_0)} dt \qquad (104)$$

$$\Delta p_{st} = -\nabla_q \int_{t_0}^{t} I^* \, dt \, , \qquad (105)$$

2) *Macroscopic local stochastic dynamics.* This case is defined by the conditions $k\Theta \neq 0$ with $\lambda_c \cup \text{ª}_L << \Delta\Omega q << \Delta\Omega_{\mathcal{L}}$ .
Given the condition $\Delta\Omega_{\mathcal{L}} >> \Delta\Omega q >> \lambda_c$ for sufficiently small $\Theta$, by (25, 27) we can set

$$< I^*, I^* >^{1/2} \cong 0 \qquad (106)$$

as well as

$$< \nabla \circ F^*, \nabla \circ F^* >^{1/2} \cong 0 \qquad (107)$$

and given the condition $\lambda_L << \Delta\Omega q << \Delta\Omega_{\mathcal{L}}$ so that it holds

$$\lim_{q \to \infty} -\nabla_q V_{qu(\mathrm{n}_0)} = 0 \qquad (108)$$

the SPDE of motion acquires the form

$$\partial_t \mathrm{n}_{(q,t)} = -\nabla_q \bullet ( \mathrm{n}_{(q,t)} \dot{q} ) + \text{¦}_{(q_\sim, t, \Theta)} \qquad (109)$$

$$< \text{¦}_{(q_\sim, t)}, \text{¦}_{(q_\sim + \text{ª}, t + \text{²})} > = \underline{\text{«}} \, \text{¢}_\sim \frac{2k\Theta}{\text{ª}_c} \text{¢}(\text{ª}) \text{¢}(\text{²}) \qquad (110)$$

$$\dot{q} = \frac{p}{m} = \nabla_q \lim_{\Delta\Omega / \text{ª}_L \to \infty} \frac{\nabla_q S}{m} = \nabla_q \{ \lim_{\Delta\Omega / \text{ª}_L \to 0} \frac{1}{m} \int_{t_0}^{t} dt( \frac{p \bullet p}{2m} + V_{(q)} - V_{qu(\mathrm{n}_0)} - I^*) \}$$

$$= \frac{1}{m} \nabla_q \{ \int_{t_0}^{t} dt \, ( \frac{p \bullet p}{2m} - V_{(q)} ) \} = \frac{p_{cl}}{m}$$

(111)

$$\dot{p}_{cl} = -\nabla_q V_{(q)}, \qquad (112)$$

# 3. Discussion

The realization of condition (89) allows fluctuations, as small as we like, to overcome the quantum force on large distance so that the quantum non-locality can only be maintained on a finite distance of order of $a_L$.

Since condition (81) is satisfied in a large number of real non-linear potentials, while the case of an infinite quantum non-locality length (such as in the linear systems) actually seems to be an exception, the universe behaves classic on its huge scale.

Generally speaking, it must be observed that even thought fluctuations are present, we may have systems characterized by an infinite quantum non-locality length $a_L$ (e.g., linear systems owing Gaussian states with *h*=2) so that, in principle, fluctuations are not sufficient to break the quantum mechanics and to lead to the classical one.

Under this light, the macro-scale description is not sufficient to obtain the classical behavior if not coupled to a finite range of quantum non-local interaction about whose the realization of (81) is a sufficient condition.

On the contrary, fluctuations may break quantum non-locality in non-linear systems (satisfying condition (81)) because, in this case, the quantum pseudo-potential decreases with distance and, beyond $a_L$, it becomes negligible with respect the fluctuations.

It must be noted that, in the large-scale description a vanishing small quantum force can be correctly neglected in presence of fluctuations but it cannot be taken out by the deterministic PDE (1) because in such a case this operation will change the structure of the equation.

This can be easily shown by noting that the presence of the QP is needed to the realization of quantum stationary states (i.e., eigenstates) that happen when the force steaming by the QP exactly balances the Hamiltonian one; if we remove the QP in (1) we also cancel the eigenstates and deeply change the structure of the QHA equation.

Moreover, the SQHA approach shows that the large-scale classical character can only emerge in the stochastic case while the deterministic classical mechanics is only a conceptual abstraction that can be achieved for $\hbar = 0$. This result, even seeming strange, is quite interesting since it can furnish the explanation why fluctuations are so wide-spread in nature and cannot be eliminated in the classical reality. The classical localization of states as a fluctuation-mediated phenomenon in non-linear systems is at glance with the current outcomes of numerical simulation on the classical to quantum transition [46].

Moreover, by observing that the QHA is constituted by two coupled first order differential equation for two real variables: the WFM and it phase and that by a substitution reduces to the Schrödinger second order differential equation of a complex variable, it follows that any solution of the Schrödinger problem is also a solution for the QHA one but not vice versa. Since in order to pass from the system of two first order differential equations of the QHA to the correspondent second order differential equation of the Schrödinger problem the wave-particle equivalence is necessary [55], this does not appear to be an important point in the deterministic limit where the wave particle equivalence holds.

On the contrary, since in presence of noise some solutions of the SQHA problem may not satisfy the wave-particle equivalence $\dot{q} = \frac{\nabla_q S_{qu}}{m}$ (i.e., for the classical states $\dot{q} = \frac{\nabla_q S_{cl}}{m}$). such SQHA states do not have their corresponding ones in the Schrödinger representation [61].

In figure 1 the traditional point of view of quantum mechanics and the SQHA one are compared.

The quantum stochastic mechanics contains as a particular case the standard quantum mechanics, the classical mechanics, with h=0, and others cases as for instance the Brownian harmonic oscillator. The SQHA model shows that the problems of the standard quantum mechanics as well as of the Brownian harmonic oscillators have their own corresponding representations [62-64] but the classical mechanics with $\hbar = 0$ has not.

Alternatively, the SQHA model shows that the stochastic classical mechanics can be achieved even if $\hbar \neq 0$.

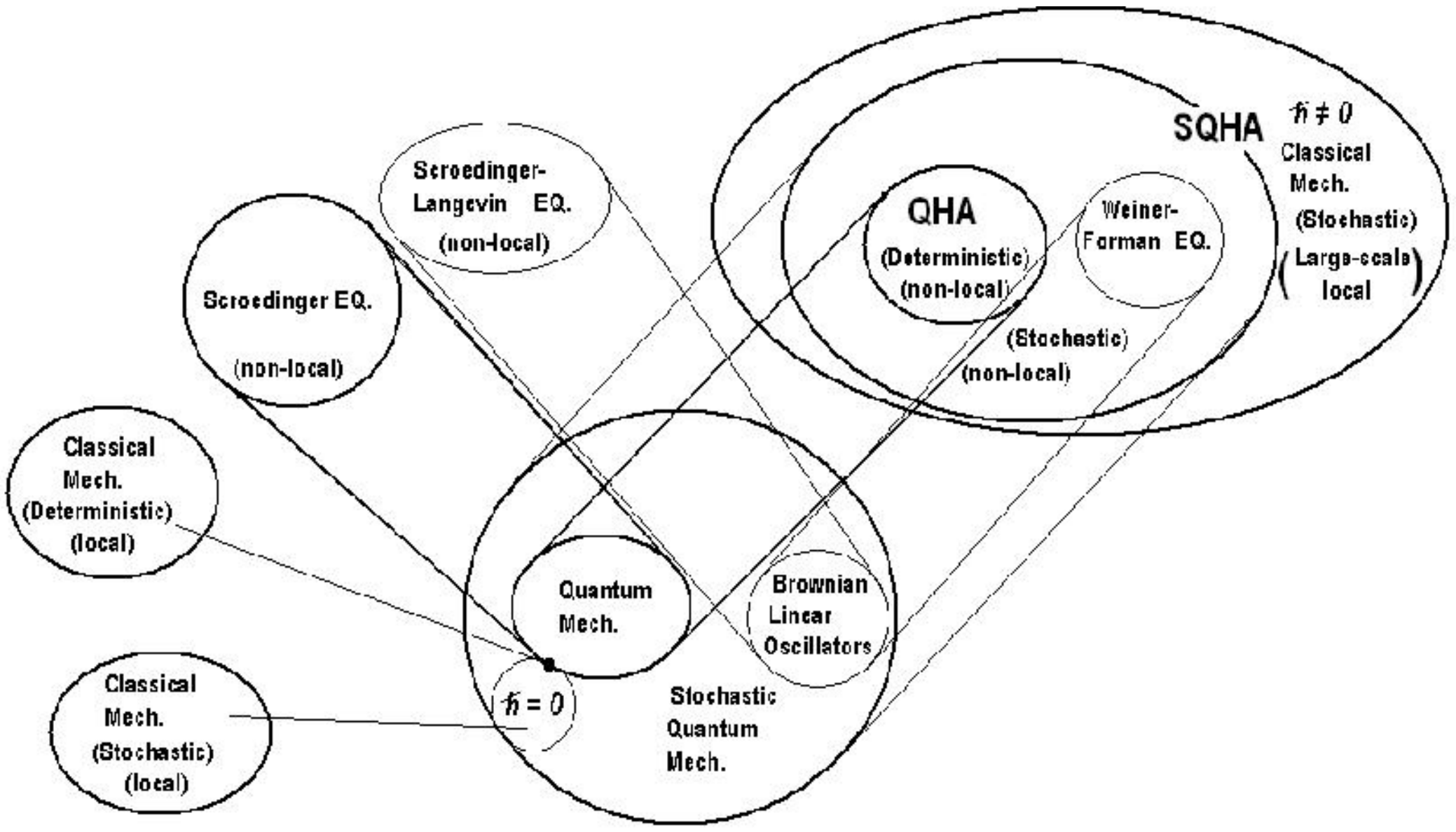

Figure 1. The correspondence among the quantum mathematical models.

Finally, it is noteworthy to note that the justification of spatially distributed noise as a consequence of an external environment or thermostat, can also seen as a consequence of the initial condition of the system. From the relativistic point of view this fact is not so unexpected since in the four dimensional space-time the initial condition (on the time co-ordinate) and the boundary conditions (on the space co-ordinates) can be treated on equal foot.
To elucidate this point, let's consider a closed system (universe) with a finite volume at the initial time. The light cones coming out from two different spatial points (not correlated at time t=0 since they are at finite distance and the light speed (propagation of interactions and information) is finite) will enlarge themselves on different spatial domains (see figure 2).

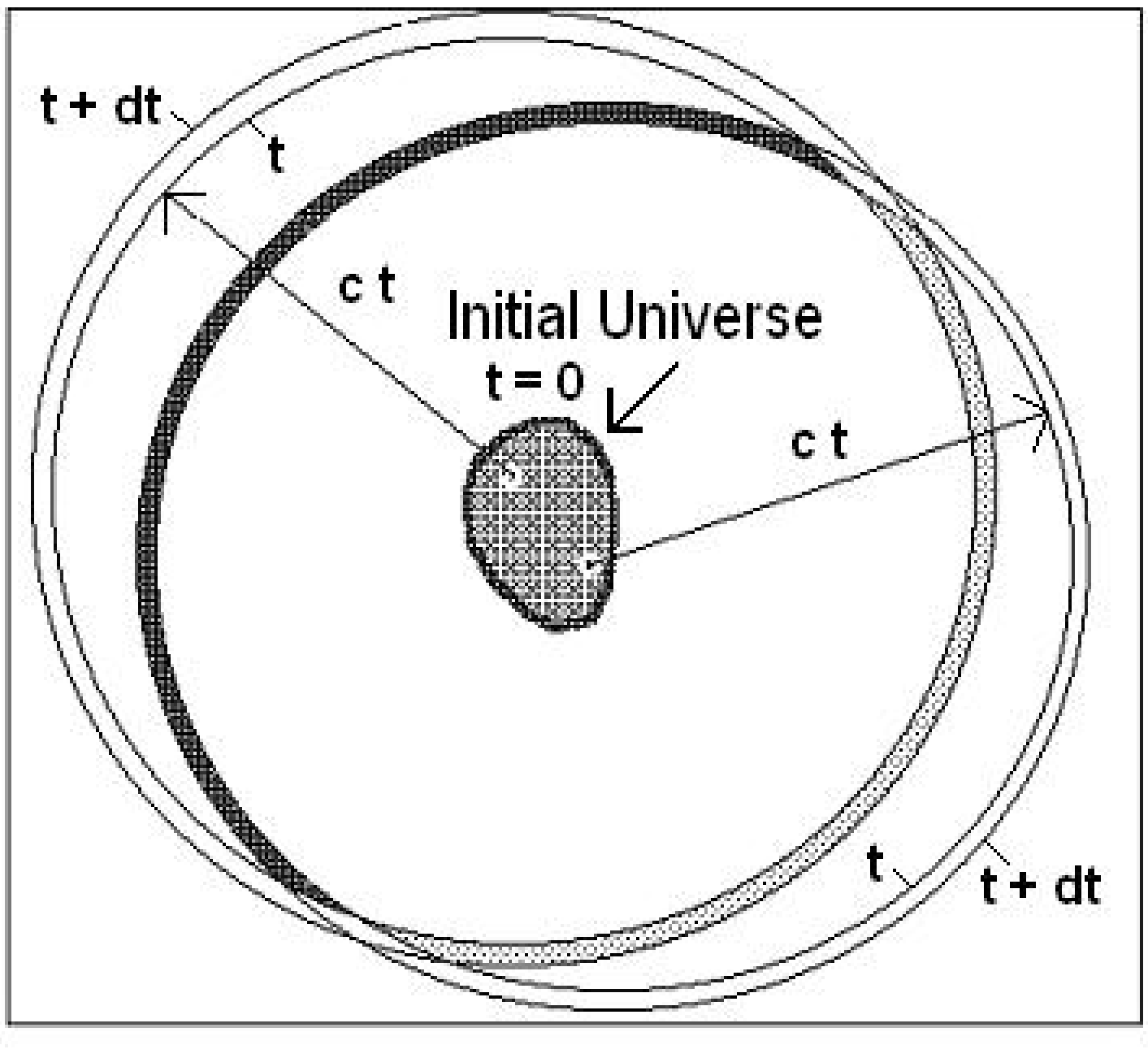

Figure 2. The expansion of cone lights from two points of a finite volume initial universe.

The interaction propagating from this two points deals with two different ensembles of vacuum oscillators. Then, after a small time increment, the vacuum oscillator that have already interacted with the force coming from one point will interact with an uncorrelated input of force coming from the other one.

Therefore, since this process is endless (in an open universe), any point of the space is subsequently reached by uncorrelated inputs coming from each point of the universe. In such a system there will always be a background of un-correlated retarded force that act as a booster.
If we acknowledge this background of infinitesimal noise as the manifestations of the lack of knowledge of the retarded effects of the interaction potentials (quantum one included) determined by the initial state of the universe (due to the finite speed of propagation of interactions together with the finite volume of the universe at the initial time) the stochastic approach (i.e., the SQHA) becomes logically self-standing and the consequence of the initial state of the universe.

# 4. Conclusion

In this paper, the standard quantum mechanics is derived as a deterministic limit of a more general stochastic QHA. The work investigate the features of the quantum behavior for a vanishing non-zero value of the fluctuations amplitude. The standard quantum mechanics in a noisy environment is always achieved on a scale much smaller than the theory-defined quantum coherence length. This is allowed by the fact that spatial fluctuations of the WFM are energetically suppressed by the quantum potential .
More analytically, the SQHA shows that in presence of spatial noise the QP modifies the shape of the fluctuations of the quantum field $\rho$ (whose spatial density in the deterministic limit represents the WFM) suppressing them on a distance much shorter than the theory-defined quantum coherence length $\lambda_c = \pi^{3/2}\hbar/(2m\, k\Theta)^{1/2}$ so that the quantum mechanics is achieved when $\lambda_c$ goes to infinity with respect to physical scale of the problem (as for the deterministic limit of null noise amplitude $\Theta = 0$).
The investigation shows that the non-local quantum interaction (that in the QHA originates by the QP) can extend itself beyond the quantum coherence length but, in the stochastic case, with a range of interaction that may be finite (maintaining the quantum non-local interactions up to a distance (the “quantum non-locality length” $\lambda_L$) whose order of magnitude can be evaluated by an integral formula. The analysis shows that the so-defined $\lambda_L$ depends by the strength of the particles interaction and by the fluctuation amplitude.
The model shows that in linear systems $\lambda_L$ can be infinite (even if $\lambda_c$ is finite) so that fluctuations are not sufficient (from the general point of view) to break the non-local interaction of the quantum mechanics.
On the contrary, for non-linear interactions, the noise may produce quantum non-locality breaking when the force of the QP decreases and becomes vanishing at large distance (beyond $\lambda_L$) being negligible with respect to the fluctuations.
The SPDE of motion exhibits various limiting dynamics, depending on the two characteristic lengths $\lambda_c$ and $\lambda_L$. For $\hbar \neq 0$ the classical stochastic behavior is achieved when $\lambda_c$ as well as $\lambda_L$ are negligibly small with respect to the physical length of the problem, while the deterministic classical mechanics is realized only for $\hbar = 0$.
The SQHA model furnishes a non-contradictory logical pathway from the quantum to the classical behavior. The quantum mechanics is deterministic while the classical one is achieved when on large distance fluctuations overcome the QP interaction (that builds up the quantum eigenstates and their superposition of states).
In the frame of the SQHA it is possible to achieve a unified understanding of quantum and classical mechanics.
The open quantum mechanics, the meso-scale quantum dynamics and the irreversible quantum phenomena are the fields where the SQHA model can be fruitful utilized and tested.
Finally, it must be underlined that the SQHA model sees the standard quantum mechanics as a deterministic “mechanics” satisfying the long waited philosophical need of such a theory.

# Appendix A

*Wave-particle equivalence in the deterministic limit of the SPDE of motion*

Here we derive the condition to which the noise must obey in the motion equation (17) in order to warrant the wave-particle equivalence in the deterministic quantum mechanics limit. In the QHA model the wave particle equivalence is warranted by the separate variable solution of type

$$^{\circ}{}_{(q,p,t)} = \mathrm{n}_{(q,t)} ¢ ( p - \nabla_q S ) \qquad \text{(A.1)}$$

where the momentum $p$ equals the gradient of the action in the quantum limit $S_{\mathrm{qu}}$ that in a plane wave represents the wave vector multiplied by the Plank's constant.
By introducing the separate variable solution $^{\circ}{}_{(q,p,t)} = \mathrm{n}_{(q,t)} ϐ_{(p,t)}$ in equation (17) we obtain the equation

$$\mathrm{n}\,\partial_t ϐ + ϐ\,\partial_t \mathrm{n} = -\mathrm{n}\nabla \bullet ( ϐ( \dot{x}_H + \dot{x}_{qu} )) - ϐ\nabla \bullet ( \mathrm{n}( \dot{x}_H + \dot{x}_{qu} )) + N_{( q,p,t,\Theta )} \qquad \text{(A.2)}$$

that is satisfied by system of equations

$$\partial_t ϐ = -\nabla \bullet ( ϐ( \dot{x}_H + \dot{x}_{qu} )) \qquad \text{(A.3)}$$

$$\partial_t \mathrm{n} = -\nabla \bullet ( \mathrm{n}( \dot{x}_H + \dot{x}_{qu} )) + ϐ^{-1} N_{( q,p,t,\Theta )} \qquad \text{(A.4)}$$

If we impose that

$$ϐ^{-1} N_{( q,p,t,\Theta )} = n( q,t,\Theta ) \qquad \text{(A.5)}$$

(A.4) is a SPDE function of q, t and Θ; while (A.3) is a deterministic PDE that can be straightforwardly solved with the general initial condition

$$ϐ_{(p,t=0)} = \prod_{\sim} ¢( p_{\sim} - p_{\sim (t=0)} ) \qquad \text{(A.6)}$$

that leads to

$$ϐ_{(p,t)} = \prod_{\sim} ¢( p_{\sim} - \nabla_q S ) \qquad \text{(A.7)}$$

where

$$S = -\int_{t_0}^{t} dt( \frac{p \bullet p}{2m} + V_{( q )} + V_{qu( \mathrm{n} )} ) = -\int_{t_0}^{t} dt( \frac{p \bullet p}{2m} + V_{( q )} + V_{qu( \mathrm{n}_0 )} + I^* ) = S_{qu} + S_{st} \qquad \text{(A.8)}$$

with

$$S_{qu} = -\int_{t_0}^{t} dt( \frac{p \bullet p}{2m} + V_{( q )} + V_{qu( \mathrm{n}_0 )} ) \qquad \text{(A.9)}$$

and hence to

$$N_{( q,p,t,\Theta )} = n( q,t,\Theta ) \prod_{\sim} ¢( p_{\sim} - \nabla_q S ) \qquad \text{(A.10)}$$

Moreover, providing, as shown in appendix D , that

$$\lim_{\Theta \to 0} I^* = 0 \qquad \text{(A.11)}$$

in the quantum deterministic limit, condition (A.7) gives

$$\lim_{\Theta \to 0} \text{ϭ}_{(p,t)} = \prod_{\sim} ¢ \left( p_{\sim} - \nabla_q S_{qu} \right) \qquad \text{(A.12)}$$

and hence that

$$\lim_{\Theta \to 0} p_{\sim} = \nabla_q S_{qu} \qquad \text{(A.13)}$$

namely, the wave-particle equivalence.

# Appendix B

*Convergence of the SQHA SPDE to the quantum deterministic limit*

The request that

$$\lim_{\Theta \to 0} < (V_{qu} - V_{qu(n_0)})^2 > = \lim_{\Theta \to 0} < I^{*2} > = 0 \qquad \text{(B.1)}$$

the root mean square of the quantum potential is vanishing in order to have the convergence to the deterministic limit for Θ tending to zero is based upon the physical requirement that the energy of the system has to be finite also in the fluctuating state.
Actually, *a priori* this is only a necessary condition while, to completely warrant the quantum deterministic limit for Θ = 0, in principle, we have to warrant (?). Given that the quantum force is the derivative of the quantum potential, a small quantum potential fluctuation may lead to a great quantum force inputs $-F^* = -\nabla_q I^*$ (that could deeply change the evolution of the system from the deterministic one). The same possibility applies also to the term $\nabla_q \bullet ( {}^\circ F^*)$. Therefore, in principle, in addition to the condition

$$\lim_{\Theta \to 0} < I^{*2} >^{1/2} = 0 \qquad \text{(B.2)}$$

we progressively must impose that

$$\lim_{\Theta \to 0} < (\nabla_q I^*)^2 >^{1/2} = 0 \qquad \text{(B.3)}$$

$$\lim_{\Theta \to 0} < (\nabla_q {}^\circ F^*)^2 >^{1/2} = 0 \qquad \text{(B.4)}$$

As shown in the section (2.1.2), in the limit of Θ going to zero, condition (B.2) is warranted by the constraint applied to the coefficients $a_0$, $a_1$, $a_2$, and $a_3$ in (54). As far as it concerns (B.3) and (B.4), since the derivative operation increases by one order the λ-exponent in (51-53), they will bring additional constraints on the coefficients $a_n$ (with j ≥ 4 by (B.3) and j ≥5 by (B.4)) in (54).
Due to the derivative structure of the quantum potential, since for Θ close to zero the terms with n≤3 predominate, condition (B.4) is less stringent than (B.3) that is less stringent than (B.2) that, hence, ultimately warrants by itself the deterministic limit.

This hierarchy shows that the energy plays a primary role also in the realization of the quantum deterministic limit.

# Appendix C

*The discretized SPDE of motion*

Practically, the introduction of condition (25) can be implemented by operating on the discrete version of the SPDE (19) whose variable reads

$$Y_{\boldsymbol{im}} = \iiint\limits_{\Delta_{\boldsymbol{i}}} d^{3n}q \iiint\limits_{\Delta_{\boldsymbol{m}}} d^{3n}p \circ_{(q,p,t)}, \qquad \text{(C.1)}$$

where the indexes *i* and *m* are actually vectors of integers

***i*** = (i $_1$ , i $_2$ , …., i$_\alpha$,…..)

***m*** = (m $_1$ , m $_2$ , …., m$_\alpha$,……)

that define the discrete point, as in the following

$$x_{\boldsymbol{im}} = (q_{\boldsymbol{i}}, p_{\boldsymbol{m}}) = (q_{i_1}, q_{i_2}, \dots, q_{i_\sim}, \dots, p_{m_1}, p_{m_2}, \dots, p_{m}\ \ , \dots) \qquad \text{(C.2)}$$

and where $\Delta_{\boldsymbol{i}}$ and $\Delta_{\boldsymbol{m}}$ are the hyper-cells sides λ and ® , respectively, of the phase-space around the discrete point x$_{\boldsymbol{im}}$ =(q$_{\boldsymbol{i}}$, p$_{\boldsymbol{m}}$) that read

$$\begin{aligned} \Delta_{\boldsymbol{i}} &= (\Delta_{i_1}, \Delta_{i_2}, \dots, \Delta_{i_\sim}, \dots) \\ \Delta_{i_\sim} &= [\,q_{i_\sim - 1}, q_{i_\sim}\,] \end{aligned} \qquad \text{(C.3)}$$

$$\begin{aligned} \Delta_{\boldsymbol{m}} &= (\Delta_{m_1}, \Delta_{m_2}, \dots, \Delta_{m_\sim}, \dots) \\ \Delta_{m_\sim} &= [\,p_{m_\sim - 1}, p_{m_\sim}\,] \end{aligned} \qquad \text{(C.4)}$$

The discretization procedure gives rise to an infinite system of discrete stochastic Langevin equations for the discrete field Y$_{\boldsymbol{i,m}}$(t). After standard manipulations [57] for Hamiltonians of type (2), the equation (19) for the discrete variable Y$_{\boldsymbol{im}}$≡ Y$_{i_\alpha\, m_\beta}$ reads

$$\frac{dY_{\boldsymbol{im}}}{dt} = -p_{m_\sim}\ \dot{Q}_{\sim \boldsymbol{ik}}\, Y_{\boldsymbol{km}} + [\,\dot{p}_\sim\ Y_{\boldsymbol{is}} - \frac{\dot{\hbar}^2}{2m} D_{qu\ \sim \boldsymbol{isjkh}}\ \mathrm{n}^0{}_{\boldsymbol{j}}\mathrm{n}^0{}_{\boldsymbol{k}}\mathrm{n}^0{}_{\boldsymbol{h}}\,]\ \dot{P}_{\sim \boldsymbol{sm}} + \dot{Y}_{qu\ \boldsymbol{im}} + N_{\boldsymbol{im}}$$

(C.5)

where

$$\dot{Q}_{\sim \boldsymbol{ik}} = (m_\sim{}^{\text{a}})^{-1}\{\,¢_{(\boldsymbol{i}+\boldsymbol{I}_{(\sim)})\boldsymbol{k}} - ¢_{\boldsymbol{ik}}\,\}, \qquad \text{(C.6)}$$

where **I**$_{(\alpha)}$ = (0,….0, 1(α-th place), 0, ….,0) and where

$$\dot{p}_\sim = \frac{\partial H}{\partial q_{i_\sim}}\ , \qquad \text{(C.7)}$$

$$n^0{}_{\boldsymbol{j}} = \sum_{\sim=1}^{3n} \sum_{m_\sim=-\infty}^{\infty} Y^0{}_{\boldsymbol{jm}} \tag{C.8}$$

$$Y^0{}_{\boldsymbol{im}} = \iiint_{\Delta_{\boldsymbol{i}}} d^{3n}q \iiint_{\Delta_{\boldsymbol{m}}} d^{3n}p\, °_{0(q,p,t)} \tag{C.9}$$

being $°_0$ the solution of the deterministic PDE and where

$$\dot{P}_{\sim \boldsymbol{sm}} = ®^{-1}\{ ¢_{\boldsymbol{s}(\boldsymbol{m}+\boldsymbol{I}_{(\sim)})} - ¢_{\boldsymbol{sm}} \}, \tag{C.10}$$

$$D_{qu\,\sim \boldsymbol{isjkh}} = a^{-3} \sum_{=0}^{3n} Y_{\boldsymbol{is}}\, ln[\frac{¢_{(\boldsymbol{i}+2\boldsymbol{I}_{(\ )}+\boldsymbol{I}_{(\sim)})\boldsymbol{j}}\, ¢_{(\boldsymbol{i}+\boldsymbol{I}_{(\sim)})\boldsymbol{k}}\, ¢_{(\boldsymbol{i}+\boldsymbol{I}_{(\ )})\boldsymbol{h}}\, n_{\boldsymbol{j}} n_{\boldsymbol{k}} n_{\boldsymbol{h}}}{¢_{(\boldsymbol{i}+2\boldsymbol{I}_{(\ )})\boldsymbol{j}}\, ¢_{(\boldsymbol{i}+\boldsymbol{I}_{(\sim)}+\boldsymbol{I}_{(\ )})\boldsymbol{k}}\, ¢_{\boldsymbol{ih}}\, n_{\boldsymbol{j}} n_{\boldsymbol{k}} n_{\boldsymbol{h}}}] \tag{C.11}$$

$$N_{\boldsymbol{im}} = \iiint_{\Delta_{\boldsymbol{i}}} d^{3n}q \iiint_{\Delta_{\boldsymbol{m}}} d^{3n}p\, ¦_{(q,t,\Theta)} ¢(p - \nabla_q S) \tag{C.12}$$

and where

$$\dot{Y}_{qu\ \boldsymbol{im}} = -\iiint_{\Delta_{\boldsymbol{i}}} d^{3n}q \iiint_{\Delta_{\boldsymbol{m}}} d^{3n}p\, \nabla \bullet ( °_{(q,p,t)} F^*) \tag{C.13}$$

represents the correction to the discretized quantum force coming from the fluctuations that critically depends by the correlation distance of the field fluctuations.

**The discrete SPDE of motion close to the deterministic limit**

In the discrete form, the small noise system of approximated equation (33-35) can be written as an infinite system Stochastic differential equations (SDE)

$$\frac{dY^0{}_{\boldsymbol{im}}}{dt} = -p_{m_\sim} \dot{Q}_{\sim \boldsymbol{ik}} Y^0{}_{\boldsymbol{km}} + [\dot{p}_\sim Y^0{}_{\boldsymbol{is}} - \frac{\dot{\hbar}^2}{2m} D_{qu\,\sim \boldsymbol{isjkh}}\, n^0{}_{\boldsymbol{j}} n^0{}_{\boldsymbol{k}} n^0{}_{\boldsymbol{h}}]\, \dot{P}_{\sim \boldsymbol{sm}} \tag{C.14}$$

$$\frac{dY^1{}_{\boldsymbol{im}}}{dt} = -p_{m_\sim} \dot{Q}_{\sim \boldsymbol{ik}} Y^1{}_{\boldsymbol{km}} + [\dot{p}_\sim Y^1{}_{\boldsymbol{is}} - \frac{\dot{\hbar}^2}{2m} D_{qu\,\sim \boldsymbol{isjkh}}\, n^0{}_{\boldsymbol{j}} n^0{}_{\boldsymbol{k}} n^0{}_{\boldsymbol{h}}]\, \dot{P}_{\sim \boldsymbol{sm}} + \dot{Y}_{qu}{}^1{}_{\boldsymbol{im}} + N_{\boldsymbol{im}} \tag{C.15}$$

$$\frac{dY^k{}_{\boldsymbol{im}}}{dt} = -p_{m_\sim} \dot{Q}_{\sim \boldsymbol{ik}} Y^k{}_{\boldsymbol{km}} + [\dot{p}_\sim Y^k{}_{\boldsymbol{is}} - \frac{\dot{\hbar}^2}{2m} D_{qu\,\sim \boldsymbol{isjkh}}\, n^0{}_{\boldsymbol{j}} n^0{}_{\boldsymbol{k}} n^0{}_{\boldsymbol{h}}]\, \dot{P}_{\sim \boldsymbol{sm}} + \dot{Y}_{qu}{}^k{}_{\boldsymbol{im}} + N_{\boldsymbol{im}} \tag{C.16}$$

where

$$\dot{Y}_{qu}{}^{k}{}_{\boldsymbol{im}} = -\iiint_{\Delta_{\boldsymbol{i}}} d^{3n}q \iiint_{\Delta_{\boldsymbol{m}}} d^{3n}p \, \nabla \bullet ( \circ_{k(q,p,t)} F^{*}{}_{k} ) \qquad \text{(C.17)}$$

and

$$Y^{k}{}_{\boldsymbol{im}} = \iiint_{\Delta_{\boldsymbol{i}}} d^{3n}q \iiint_{\Delta_{\boldsymbol{m}}} d^{3n}p \circ_{k(q,p,t)} \qquad \text{(C.18)}$$

In the case of $\Theta$ converging to zero, the small noise series expansion

$$Y^{k}{}_{\boldsymbol{im}} = Y^{k}{}_{\boldsymbol{im}\,0} + Y^{k}{}_{\boldsymbol{im}\,1} + \ldots + Y^{k}{}_{\boldsymbol{im}\,u} + \ldots \qquad \text{(C.19)}$$

can be used to solve the system of equation (C.14-C.16) . In this case, the first order solution

$$Y^{1}{}_{\boldsymbol{im}} = Y^{1}{}_{\boldsymbol{im}\,0} + Y^{1}{}_{\boldsymbol{im}\,1} + \ldots + Y^{1}{}_{\boldsymbol{im}\,u} + \ldots = Y^{0}{}_{\boldsymbol{im}} + Y^{1}{}_{\boldsymbol{im}\,1} + \ldots + Y^{1}{}_{\boldsymbol{im}\,u} + \ldots \qquad \text{(C.20)}$$

where it has been introduced the information that the zero order solution $Y^{1}{}_{\boldsymbol{im}\,0}$ is given by the deterministic one $Y^{0}{}_{\boldsymbol{im}}$ , can be used to solve (C.15) leading to

$$dY^{1}{}_{\boldsymbol{im}} \cong dY^{0}{}_{\boldsymbol{im}} + dY^{1}{}_{\boldsymbol{im}\,1} \,. \qquad \text{(C.21)}$$

Moreover, being

$$\dot{Y}_{qu}{}^{1}{}_{\boldsymbol{im}} = 0 \,, \qquad \text{(C.22)}$$

the SDE (C.15) [57] at the first order reads

$$dY^{1}{}_{\boldsymbol{im}\,1} = -K_{(Y^{0}{}_{\boldsymbol{im}})} Y^{1}{}_{\boldsymbol{im}\,1} dt + N_{\boldsymbol{im}} dt \qquad \text{(C.23)}$$

where

$$K_{(Y^{0}{}_{\boldsymbol{im}})} = -\{ \frac{\partial \{ p_{m\sim} \dot{Q}_{\sim\boldsymbol{ik}} Y^{0}{}_{\boldsymbol{km}} + [ \dot{p}_{\sim} Y^{0}{}_{\boldsymbol{is}} - \frac{\dot{\hbar}^{2}}{2m} D_{qu\,\sim\boldsymbol{isjkh}} \, \mathrm{n}^{0}{}_{\boldsymbol{j}} \mathrm{n}^{0}{}_{\boldsymbol{k}} \mathrm{n}^{0}{}_{\boldsymbol{h}} ] \dot{P}_{\sim\boldsymbol{sm}} \}}{\partial Y^{0}{}_{\boldsymbol{im}}} \} \qquad \text{(C.24)}$$

that with the initial conditions descending from (30) that reads $Y^{1}{}_{\boldsymbol{im}\,1(t=0)} = 0$ , leads to

$$dY^{1}{}_{\boldsymbol{im}\,1} = N_{\boldsymbol{im}} dt \qquad \text{(C.25)}$$

and thence to the first order WFM fluctuations [57]

$$dY^{1}{}_{\boldsymbol{im}} \cong dY^{0}{}_{\boldsymbol{im}} + dY^{1}{}_{\boldsymbol{im}\,1}$$
$$= dY^{0}{}_{\boldsymbol{im}} + \{ \iiint_{\Delta_{\boldsymbol{i}}} d^{3n}q \iiint_{\Delta_{\boldsymbol{m}}} d^{3n}p \,¦_{(q,t,\Theta)} ¢(p - \nabla_q S)\} dt \qquad \text{(C.26)}$$

Moreover, being

$$\text{n}^{\text{k}}{}_{\boldsymbol{j}} = \sum_{\sim=1}^{3n} \sum_{m_{\sim}=-\infty}^{\infty} Y^{\text{k}}{}_{\boldsymbol{jm}} \qquad \text{(C.27)}$$

in the small noise approximation it follows that

$$\text{n}^{1}{}_{\boldsymbol{i}} \cong \text{n}^{0}{}_{\boldsymbol{i}} + \text{n}^{1}{}_{\boldsymbol{i}\,1} \qquad \text{(C.28)}$$

$$d\text{n}^{1}{}_{\boldsymbol{i}} \cong d\text{n}^{0}{}_{\boldsymbol{i}} + d\text{n}^{1}{}_{\boldsymbol{i}\,1}$$
$$= d\text{n}^{0}{}_{\boldsymbol{i}} + \{ \iiint_{\Delta_{\boldsymbol{i}}} d^{3n}q \,¦_{(q,t,\Theta)} \sum_{\sim=1}^{3n} \sum_{m_{\sim}=-\infty}^{\infty} \iiint_{\Delta_{\boldsymbol{m}}} d^{3n}p \,¢(p - \nabla_q S)\} dt \qquad \text{(C.29)}$$

that passing to the continuous limit (i.e.,

$$\sum_{\sim=1}^{3n} \sum_{m_{\sim}=-\infty}^{\infty} \iiint_{\Delta_{\boldsymbol{m}}} d^{3n}p \,¢(p - \nabla_q S)$$
$$= \sum_{\sim=1}^{3n} \sum_{m_{\sim}=-\infty}^{\infty} ¢(p_{\boldsymbol{m}} - \nabla_q S) \to \iiint d^{3n}p ¢(p - \nabla_q S) = 1 \qquad \text{(C.30)}$$

gives

$$\frac{d\text{n}_{(q)}}{dt} \cong \frac{d\text{n}^{1}{}_{(q)}}{dt} \cong \frac{d\text{n}^{0}{}_{(q)}}{dt} + \frac{d\text{n}^{1}{}_{1(q)}}{dt} = \frac{d\text{n}^{0}{}_{(q)}}{dt} + ¦_{(q,t,\Theta)} \qquad \text{(C.31)}$$

# Appendix D

*Quantum potential fluctuations in the small noise limit*

Given that

$$\lim_{\Theta \to 0} \langle (V_{qu} - V_{qu(\text{n}_0)})^2 \rangle = \lim_{\Theta \to 0} \{ \langle V_{qu}{}^2 \rangle - 2 \langle V_{qu} V_{qu(\text{n}_0)} \rangle + \langle V_{qu(\text{n}_0)} \rangle^2 \}$$

(D.1)

and that

$$\lim_{\Theta \to 0} \langle V_{qu}, V_{qu} \rangle = \lim_{\Theta \to 0} \{ \langle V_{qu}{}^2 \rangle - \langle V_{qu} \rangle^2 \} \qquad \text{(D.2)}$$

it clearly appears that the QP divergence generated by the shape of the WFM fluctuations in both terms (D.1,D.2) are brought by the quadratic mean $\lim_{\Theta \to 0} < V_{qu}{}^{2} >$ that is sensitive to amplitude of the QP fluctuations.

On the contrary, the linear mean terms: $< V_{qu} V_{qu(\text{n}_0)} >$ and $< V_{qu} >$ are expected to be quite insensitive to fluctuations since in the small noise approximation the probability transition function is Gaussian and fluctuations are practically symmetric respect the change of sign. Therefore, both $\lim_{\Theta \to 0} < V_{qu}, V_{qu} >$ and $\lim_{\Theta \to 0} < (V_{qu} - V_{qu(\text{n}_0)})^2 >$ do not diverge (as a function of the shape of the WFM fluctuations) if $\lim_{\Theta \to 0} < V_{qu}{}^{2} >$ does not, and vice versa.

Thence, in order to warrant (D.1) we impose that $\lim_{\Theta \to 0} < V_{qu}, V_{qu} >$ does not diverge in order to derive the correlation length of the WFM fluctuations.

Given the quantum potential

$$V_{qu} = -(\frac{\hbar^2}{2m})\{ \text{n}^{-1}(\nabla_q \bullet \nabla_q \text{n}) - \text{n}^{-2}(\nabla_q \text{n} \bullet \nabla_q \text{n}) \} \qquad \text{(D.3)}$$

the divergence of $\lim_{\Theta \to 0} < V_{qu}, V_{qu} >$ due to the shape of WFM fluctuations can be evaluated by writing down the discrete expression of the derivative terms as follows

$$\nabla_q \text{n} = (\frac{\partial \text{n}}{\partial q_1}, ......, \frac{\partial \text{n}}{\partial q_{3n}},)$$
$$\frac{\partial \text{n}}{\partial q_{\sim}} = \lim_{\text{a} \to 0} \text{a}^{-1}\{ \text{n}_{(q_{\sim(i+1)},t)} - \text{n}_{(q_{\sim(i)},t)} \} \qquad \text{(D.4)}$$

$$\nabla_q \text{n} \bullet \nabla_q \text{n} = \lim_{\text{a} \to 0} \sum_{\sim} \text{a}^{-2} [\text{n}_{(q_{\sim(i+1)}} - \text{n}_{(q_{\sim(i)}}]^2 \qquad \text{(D.5)}$$

and

$$\nabla_q \bullet \nabla_q \text{n} = \lim_{\text{a} \to 0} \sum_{\sim} \text{a}^{-2} [\text{n}_{(q_{\sim(i+2)}} - 2\text{n}_{(q_{\sim(i+1)}} + \text{n}_{(q_{\sim(i)}}], \qquad \text{(D.6)}$$

by deriving $\text{n}_{(q_{\sim(i)}}$ from (C.31) (once (74*) by using (C.27) and the correlation function (40) we obtain

$$< \text{n}_{(q_{\sim},t)}, \text{n}_{(q\ +\text{a},t)} > \cong < \text{n}^1{}_{(q_{\sim},t)}, \text{n}^1{}_{(q_{\sim(i)}+\text{a},t)} > = < \text{n}^0 + \int_0^t ¦_{(q_{\sim},t)} dt, \text{n}^0 + \int_0^t ¦_{(q\ +\text{a},t)} dt >$$
$$= < \text{n}^0{}_{(q_{\sim},t)}, \text{n}^0{}_{(q\ +\text{a},t)} > + < \int_0^t ¦_{(q_{\sim},t)} dt, \int_0^t ¦_{(q\ +\text{a},t)} dt > \cong < \text{n}^0, \text{n}^0 > + g(0) G(\text{a}) ¢_{\sim}\ \Delta t \qquad \text{(D.7)}$$

that for normalized states (localized particles) for which it holds

$$< \text{n}^0, \text{n}^0 > = 0 \qquad \text{(D.8)}$$

since $\text{n}^0$ is a deterministic solution, finally reads

$$< n_{(q_{\sim},t)}, n_{(q \; + a,t)} > \cong g(0) G(a) ¢_{\sim} \; \Delta t \qquad \text{(D.9)}$$

where the relations

$$< n^{1}{}_{1(q_{\sim},t)}, n^{1}{}_{1(q_{\sim(i)} + a,t)} > = < \int_{t}^{t+\Delta t} ¦_{(q_{\sim},t)} dt, \int_{t}^{t+\Delta t} ¦_{(q \; + a,t)} dt > = g(0) G(a) ¢_{\sim} \; \Delta t \qquad \text{(D.10)}$$

$$n_{(q,t)} \cong n^{0}{}_{(q,t)} + n^{1}{}_{1}(q,t) = n^{0}{}_{(q,t)} + \int_{0}^{t} ¦_{(q,t,\Theta)} dt \qquad \text{(D.11)}$$

have been used.

Therefore, we can write

$$\begin{aligned}
< \frac{\partial n}{\partial q_{\sim}}, \frac{\partial n}{\partial q} > = & \lim_{a \to 0} a^{-2} \{ < ( n_{(q_{\sim(i+1)},t)} - n_{(q_{\sim(i)},t)} )( n_{(q \; (i+1),t)} - n_{(q \; (i),t)} ) > \\
& - < ( n_{(q_{\sim(i+1)},t)} - n_{(q_{\sim(i)},t)} ) >< ( n_{(q \; (i+1),t)} - n_{(q \; (i),t)} ) > \} \\
= & \lim_{a \to 0} a^{-2} \{ < ( n_{(q_{\sim(i+1)},t)}, n_{(q \; (i+1),t)} ) > + < ( n_{(q_{\sim(i)},t)}, n_{(q \; (i),t)} ) > \\
& - < ( n_{(q_{\sim(i+1)},t)}, n_{(q \; (i),t)} ) > - < ( n_{(q_{\sim(i)},t)}, n_{(q \; (i+1),t)} ) \}
\end{aligned} \qquad \text{(D.12)}$$

Moreover, since for uncorrelated noises on different spatial components we have

$$< ( n_{(q_{\sim},t)}, n_{(q \;\; ,t)} ) > = ¢_{\sim} \;\; < ( n_{(q_{\sim},t)}, n_{(q_{\sim},t)} ) > \qquad \text{(D.13)}$$

it follows that

$$\begin{aligned}
\sum_{\sim} < \frac{\partial n}{\partial q_{\sim}}, \frac{\partial n}{\partial q} > = & \lim_{a \to 0} a^{-2} \sum_{\sim} ¢_{\sim} \;\; \{ < ( n_{(q_{\sim(i+1)},t)}, n_{(q \; (i+1),t)} ) > + < ( n_{(q_{\sim(i)},t)}, n_{(q \; (i),t)} ) > \\
& - < ( n_{(q_{\sim(i+1)},t)}, n_{(q \; (i),t)} ) > - < ( n_{(q_{\sim(i)},t)}, n_{(q \; (i+1),t)} ) \} \\
= & \lim_{a \to 0} a^{-2} \sum_{\sim} \{ < ( n_{(q_{\sim(i+1)},t)}, n_{(q_{\sim(i+1)},t)} ) > + < ( n_{(q_{\sim(i)},t)}, n_{(q_{\sim(i)},t)} ) > \\
& - < ( n_{(q_{\sim(i+1)},t)}, n_{(q_{\sim(i)},t)} ) > - < ( n_{(q_{\sim(i)},t)}, n_{(q_{\sim(i+1)},t)} ) \} \\
= & \lim_{a \to 0} a^{-2} \sum_{\sim} 2 < g_{\sim(i+1)} > [1 - G_{\sim}(a)]
\end{aligned} \qquad \text{(D.14)}$$

where

$$< g_{\sim(i+1)} > = \frac{g_{\sim(i)} + g_{\sim(i+1)}}{2} \qquad \text{(D.15)}$$

$$g_{\sim(i)} = < ( n_{(q_{\sim(i)},t)}, n_{(q_{\sim(i)},t)} ) > = < n^{0}, n^{0} > + g(0) \Delta t \cong g(0) \Delta t \qquad \text{(D.16)}$$

$$G_{\sim}(\text{a}) = \frac{<(\text{n}_{(q_{\sim(i+1)},t)}, \text{n}_{(q_{\sim(i)},t)})>}{<g_{\sim(i+1)}>} \tag{D.17}$$

Given that the WFM fluctuations at lowest order are approximately Gaussian, all higher moments are functions of the moments of second order. By using the identity holding for a Gaussian random variables $\Gamma = (\Gamma_{\sim}\ )$

$$\begin{aligned} <\Gamma \bullet \Gamma, \Gamma \bullet \Gamma> &=< \Gamma_{\sim} \Gamma_{\sim}, \Gamma\ \ \Gamma\ \ >= \sum_{\sim} < \Gamma_{\sim}{}^{2}, \Gamma\ \ {}^{2} > \\ &= \sum_{\sim} [< \Gamma_{\sim}, \Gamma\ \ > + < \Gamma_{\sim} >< \Gamma\ \ >]^{2} \end{aligned} \tag{D.18}$$

we obtain

$$< \nabla_q \text{n} \bullet \nabla_q \text{n}, \nabla_q \text{n} \bullet \nabla_q \text{n} > = 2\sum_{\sim} [< \frac{\partial \text{n}}{\partial q_{\sim}}, \frac{\partial \text{n}}{\partial q} > + < \frac{\partial \text{n}}{\partial q_{\sim}} >< \frac{\partial \text{n}}{\partial q} >]^{2} \tag{D.19}$$

Since the terms $< \frac{\partial \text{n}}{\partial q_{\sim}} >$ reads

$$\begin{aligned} < \frac{\partial \text{n}}{\partial q_{\sim}} > &=< \frac{\partial(\text{n}^{0} + \int_{0}^{t} ¦_{(q_{\sim},t)} dt)}{\partial q_{\sim}} >=< \frac{\partial \text{n}^{0}}{\partial q_{\sim}} + \frac{\partial \int_{0}^{t} ¦_{(q_{\sim},t)} dt}{\partial q_{\sim}} >=< \frac{\partial \text{n}^{0}}{\partial q_{\sim}} > + < \int_{0}^{t} \frac{\partial ¦_{(q_{\sim},t)}}{\partial q_{\sim}} dt > \\ &=< \frac{\partial \text{n}^{0}}{\partial q_{\sim}} > + \int_{0}^{t} < \frac{\partial ¦_{(q_{\sim},t)}}{\partial q_{\sim}} > d =< \frac{\partial \text{n}^{0}}{\partial q_{\sim}} > + \int_{0}^{t} < \lim_{\text{a} \to 0} \frac{¦_{(q_{\sim}+\text{a},t)} - ¦_{(q_{\sim},t)}}{\text{a}} > dt \\ &=< \frac{\partial \text{n}^{0}}{\partial q_{\sim}} > + \lim_{\text{a} \to 0} \text{a}^{-1} \int_{0}^{t} (< ¦_{(q_{\sim}+\text{a},t)} > - < ¦_{(q_{\sim},t)} >) dt =< \frac{\partial \text{n}^{0}}{\partial q_{\sim}} > \end{aligned} \tag{D.20}$$

we obtain

$$< \nabla_q \text{n} \bullet \nabla_q \text{n}, \nabla_q \text{n} \bullet \nabla_q \text{n} > = 2\sum_{\sim} [< \frac{\partial \text{n}}{\partial q_{\sim}}, \frac{\partial \text{n}}{\partial q} > + < \frac{\partial \text{n}^{0}}{\partial q_{\sim}} >< \frac{\partial \text{n}^{0}}{\partial q} >]^{2} \tag{D.21}$$

Given that the terms $< \frac{\partial \text{n}^{0}}{\partial q_{\sim}} >$ (that are independent by the fluctuations) are ineffective for the divergence of the root mean square QP fluctuations, we can discharge it to obtain

$$< \nabla_q \text{n} \bullet \nabla_q \text{n}, \nabla_q \text{n} \bullet \nabla_q \text{n} > = 2\sum_{\sim} [< \frac{\partial \text{n}}{\partial q_{\sim}}, \frac{\partial \text{n}}{\partial q} >^{2} + 2A < \frac{\partial \text{n}}{\partial q_{\sim}}, \frac{\partial \text{n}}{\partial q} >] \tag{D.22}$$

with

$$A =< \frac{\partial n^0}{\partial q_\sim} >< \frac{\partial n^0}{\partial q} > \qquad \text{(D.23)}$$

Therefore, at the smallest order in Θ, we obtain

$$< \nabla_q n \bullet \nabla_q n, \nabla_q n \bullet \nabla_q n > = \lim_{ª \to 0} ª^{-4} \{ \sum_{\sim} 4 < g_{\sim (i+1)} >^2 [1 - G_\sim(ª)] \}^2 + \lim_{ª \to 0} ª^{-2} \sum_{\sim} 4A < g_{\sim (i+1)} > [1 - G_\sim(ª)]$$

(D.24)

where

$$\lim_{ª \to 0} G_\sim(ª) = \lim_{ª \to 0} \frac{<(n(q_{\sim (i+1)},t), n(q_{\sim (i)},t))>}{< g_{\sim (i+1)} >} = \frac{<(n(q_\sim + ª, t), n(q_\sim, t))>}{g(0)t} = G(ª)$$

(D.25)

Moreover, by using the identity

$$\lim_{ª \to 0} < g_{\sim (i+1)} > =< n^0, n^0 > + g(0)\Delta t = g(0)\Delta t$$

(D.26)

finally (D.24) reads

$$< \nabla_q n \bullet \nabla_q n, \nabla_q n \bullet \nabla_q n > = \lim_{ª \to 0} ª^{-4} \{ \sum_{\sim} 4\Delta t^2 \, g(0)^2 [1 - G(ª)] \}^2 + \lim_{ª \to 0} ª^{-2} \sum_{\sim} 4A\Delta t \, g(0)[1 - G_\sim(ª)]$$

(D.27)

Furthermore, writing down the second partial derivative in the discrete form as

$$\nabla_q \bullet \nabla_q n = \lim_{ª \to 0} ª^{-2} \{ n(q_{\sim (i+2)},t) - 2n(q_{\sim (i+1)},t) + n(q_{\sim (i)},t) \} \qquad \text{(D.28)}$$

applying the condition of uncorrelated WFM fluctuations on different spatial components, the variance of (D.28) reads

$$\begin{aligned} < \nabla_q \bullet \nabla_q n, \nabla_q \bullet \nabla_q n > &= \lim_{ª \to 0} ª^{-4} \{ < \{ \sum_{\sim} n(q_{\sim (i+2)},t) - 2n(q_{\sim (i+1)},t) + n(q_{\sim (i)},t) \}^2 > \\ &\quad - \sum_{\sim} < n(q_{\sim (i+2)},t) - 2n(q_{\sim (i+1)},t) + n(q_{\sim (i)},t) >^2 \} \\ &= \lim_{ª \to 0} ª^{-4} \sum_{\sim} \{ < \{ n(q_{\sim (i+2)},t) - 2n(q_{\sim (i+1)},t) + n(q_{\sim (i)},t) \}^2 > \\ &\quad - < n(q_{\sim (i+2)},t) - 2n(q_{\sim (i+1)},t) + n(q_{\sim (i)},t) >^2 \} \end{aligned}$$

(D.29)

that unwinding the terms inside (D.29), leads to

$$< \nabla_q \bullet \nabla_q \nu, \nabla_q \bullet \nabla_q \nu >=$$

$$= \lim_{a \to 0} a^{-4} \sum_{\sim} \{< \nu_{(q_{\sim(i+2)},t)}, \nu_{(q_{\sim(i+2)},t)} > +4 < \nu_{(q_{\sim(i+1)},t)}, \nu_{(q_{\sim(i+1)},t)} > + < \nu_{(q_{\sim(i)},t)}, \nu_{(q_{\sim(i)},t)} >$$

$$+ 2 < \nu_{(q_{\sim(i+2)},t)}, \nu_{(q_{\sim(i)},t)} > -4 < \nu_{(q_{\sim(i+2)},t)}, \nu_{(q_{\sim(i+1)},t)} > -4 < \nu_{(q_{\sim(i+1)},t)}, \nu_{(q_{\sim(i)},t)} > \}$$

$$= \lim_{a \to 0} a^{-4} \sum_{\sim} 2 < g_{\sim(i+2)} > [1 + \frac{G_{\sim}(2a)}{3} - \frac{4G_{\sim}(a) < g_{\sim(i+1)} >}{< g_{\sim(i+2)} >}]$$

(D.30)

where

$$< g_{\sim(i+2)} >= \frac{g_{\sim(i)} + 4g_{\sim(i+1)} + g_{\sim(i+2)}}{2} \quad (D.31)$$

$$G_{\sim}(2a) = \frac{3 < (\nu_{(q_{\sim(i+2)},t)}, \nu_{(q_{\sim(i)},t)}) >}{< g_{\sim(i+2)} >} \quad (D.32)$$

(10.19)

Moreover, by using the identities

$$\lim_{a \to 0} < g_{\sim(i+2)} >= \lim_{a \to 0} \frac{g_{\sim(i)} + 4g_{\sim(i+1)} + g_{\sim(i+2)}}{2}$$

$$\cong \frac{g(0)\Delta t + 4g(0)\Delta t + g(0)\Delta t}{2} \cong 3g(0)\Delta t$$

(D.33)

$$\lim_{a \to 0} G_{\sim}(2a) = \lim_{a \to 0} \frac{3 < (\nu_{(q_{\sim(i+2)},t)}, \nu_{(q_{\sim(i)},t)}) >}{< g_{\sim(i+2)} >} = \frac{< (\nu_{(q_{\sim}+2a,t)}, \nu_{(q_{\sim},t)}) >}{g(0)t} = G(2a)$$

(D.34)

it follows that

$$< \nabla_q \bullet \nabla_q \nu, \nabla_q \bullet \nabla_q \nu >= \lim_{a \to 0} a^{-4} \sum_{\sim} 2\Delta t\, g(0)[3 + G(2a) - 4G(a)]$$

(D.35)

Furthermore, being that

$$< \nabla_q \bullet \nabla_q \nu, \nabla_q \nu \bullet \nabla_q \nu >=$$

$$= \lim_{a \to 0} a^{-4} < \{\sum_{\sim} \nu_{(q_{\sim(i+2)},t)} - 2\nu_{(q_{\sim(i+1)},t)} + \nu_{(q_{\sim(i)},t)}\}$$

$$\{\sum_{\sim} \nu^2_{(q_{\sim(i+2)},t)} - 2\nu_{(q_{\sim(i+1)},t)}\nu_{(q_{\sim(i)},t)} + \nu^2_{(q_{\sim(i)},t)}\} > -$$

$$- \sum_{\sim} < \nu_{(q_{\sim(i+2)},t)} - 2\nu_{(q_{\sim(i+1)},t)} + \nu_{(q_{\sim(i)},t)} >$$

$$< \{\sum_{\sim} \nu^2_{(q_{\sim(i+2)},t)} - 2\nu_{(q_{\sim(i+1)},t)}\nu_{(q_{\sim(i)},t)} + \nu^2_{(q_{\sim(i)},t)}\} > \}$$

(D.36)

by grouping the terms and by using the property that the third moments of a Gaussian random variable are null, it follows that

$$
\begin{aligned}
&< \nabla_q \bullet \nabla_q \mathrm{n}, \nabla_q \mathrm{n} \bullet \nabla_q \mathrm{n} >= \\
&= \lim_{\mathrm{a}\to 0} \mathrm{a}^{-4} \sum_{\sim} \{ < \mathrm{n}_{(q_{\sim(i+2)},t)}, \mathrm{n}^2_{(q_{\sim(i+1)},t)} > -2 < \mathrm{n}_{(q_{\sim(i+1)},t)}, \mathrm{n}^2_{(q_{\sim(i+1)},t)} > \\
&+ < \mathrm{n}_{(q_{\sim(i)},t)}, \mathrm{n}^2_{(q_{\sim(i+1)},t)} > + < \mathrm{n}_{(q_{\sim(i+2)},t)}, \mathrm{n}^2_{(q_{\sim(i)},t)} > -2 < \mathrm{n}_{(q_{\sim(i+1)},t)}, \mathrm{n}^2_{(q_{\sim(i)},t)} > \\
&+ < \mathrm{n}_{(q_{\sim(i)},t)}, \mathrm{n}^2_{(q_{\sim(i)},t)} > -2 < \mathrm{n}_{(q_{\sim(i+1)},t)} \mathrm{n}_{(q_{\sim(i)},t)}, \mathrm{n}_{(q_{\sim(i+2)},t)} > \\
&\qquad + 4 < \mathrm{n}_{(q_{\sim(i+1)},t)} \mathrm{n}_{(q_{\sim(i)},t)}, \mathrm{n}_{(q_{\sim(i+1)},t)} > \} = 0
\end{aligned}
$$

(D.38)

Therefore, in force of the above calculations, finally, the quantum potential variance reads

$$
\begin{aligned}
&\lim_{\mathrm{a}\to 0} < V_{qu}, V_{qu} > \\
&= \lim_{\mathrm{a}\to 0} \left(\frac{\hbar^2}{2m}\right)^2 [\mathrm{d}_1^{-2} < \nabla_q \bullet \nabla_q \mathrm{n}, \nabla_q \bullet \nabla_q \mathrm{n} > \text{-} \mathrm{d}_2^{-4} < \nabla_q \mathrm{n} \bullet \nabla_q \mathrm{n}, \nabla_q \mathrm{n} \bullet \nabla_q \mathrm{n} > ]
\end{aligned}
$$

(D.39)

that by (D.27,D.35) leads to

$$
\begin{aligned}
\lim_{\mathrm{a}\to 0} < V_{qu}, V_{qu} > &= 2\left(\frac{\hbar^2}{2m}\right)^2 \lim_{\mathrm{a}\to 0} \{ \mathrm{d}_1^{-2} g(0) \Delta t \sum_{\sim} \mathrm{a}^{-4} [3 + G(2\mathrm{a}) - 4G(\mathrm{a})] \\
&- \mathrm{d}_2^{-4} [ g(0)^2 \, 4\Delta t^2 (\sum_{\sim} \mathrm{a}^{-2} [1 - G(\mathrm{a})])^2 + g(0) 4A\Delta t \sum_{\sim} \mathrm{a}^{-2} [1 - G_{\sim}(\mathrm{a})] \}
\end{aligned}
$$

(D.40)

where the (weighted) mean particle densities $\mathrm{d}_1$, $\mathrm{d}_2$, given by (??) are finite positive values (e.g. for n constant over all the space $\mathrm{d}_1 = \mathrm{d}_2 = \mathrm{n}$)

Moreover, given that

$$\lim_{\mathrm{a}\to 0} \sum_{\sim} \mathrm{a}^{-2} [1 - G_{\sim}(\mathrm{a})] = -\frac{a_2}{\mathrm{a}_c^{\,2}} \qquad \text{(D.41)}$$

$$\lim_{\mathrm{a}\to 0} \sum_{\sim} \mathrm{a}^{-4} [1 - G_{(\mathrm{a})}]^2 = \frac{a_2^{\,2}}{\mathrm{a}_c^{\,4}} \qquad \text{(D.42)}$$

$$\lim_{\mathrm{a}\to 0} \sum_{\sim} \mathrm{a}^{-4} [3 + G_{(2\mathrm{a})} - 4G_{(\mathrm{a})}] = \frac{16 a_4}{\mathrm{a}_c^{\,4}} \qquad \text{(D.43)}$$

it follows that, at smallest order in kΘ, $< V_{qu}, V_{qu} >$ goes like

$$\lim_{\mathrm{a}\to 0} < V_{qu}, V_{qu} >= \lim_{\mathrm{a}\to 0} < I^*, I^* > \propto \frac{g(0)}{\mathrm{a}_c^{\,2}} \propto (k\Theta)^3 \qquad \text{(D.44)}$$

Moreover, given that

$$\lim_{a\to 0} < I^* > \approx 0$$

it also follows that

$$\lim_{a\to 0} < I^*,I^* > = \lim_{a\to 0} < I^{*2} > \propto \frac{g(0)}{a_c^{\;2}} \propto (k\Theta)^3 .$$

Furthermore, given also that

$$\nabla I^* \approx \frac{< I^{*2} >^{\frac{1}{2}}}{a_c}$$

$$\lim_{a\to 0} < (\nabla I^*)^2 > = \lim_{a\to 0} \frac{< I^{*2} >}{a_c^{\;2}} = \propto \frac{g(0)}{a_c^{\;4}} \propto (k\Theta)^4$$

and that

$$\lim_{\Theta\to 0} < (\nabla_q \circ F^*)^2 > \; = \lim_{\Theta\to 0} < ( \circ\nabla_q\nabla_q I^* + \nabla_q I^* \nabla_q \circ )^2 >$$

$$= \lim_{\Theta\to 0} < ( \circ\nabla_q\nabla_q I^*)^2 > + < (\nabla_q I^* \nabla_q \circ )^2 > +2 < \circ\nabla_q\nabla_q I^* \nabla_q I^* \nabla_q \circ >$$

$$\approx \lim_{\Theta\to 0} < \circ \frac{< \nabla I^{*2} >}{a_c^{\;2}} > + < \frac{< I^{*2} >}{a_c^{\;2}} (\nabla_q \circ )^2 > +2 < \circ \frac{< \nabla I^{*2} >^{\frac{1}{2}}}{a_c} \frac{< I^{*2} >^{\frac{1}{2}}}{a_c} \nabla_q \circ > \approx$$

$$\approx \lim_{\Theta\to 0} < \circ \frac{< I^{*2} >}{a_c^{\;4}} > + < \frac{< I^{*2} >}{a_c^{\;2}} (\nabla_q \circ )^2 > +2 < \circ \frac{< I^{*2} >^{\frac{1}{2}}}{a_c^{\;2}} \frac{< I^{*2} >^{\frac{1}{2}}}{a_c} \nabla_q \circ > \approx$$

$$\approx \lim_{\Theta\to 0} \propto < \circ \frac{(k\Theta)^3}{a_c^{\;4}} > + \propto < \frac{(k\Theta)^3}{a_c^{\;2}} (\nabla_q \circ )^2 > + \propto < \circ \frac{(k\Theta)^3}{a_c^{\;3}} \nabla_q \circ > \approx$$

$$\approx \lim_{\Theta\to 0} \propto (k\Theta)^5 < \circ > + \propto (k\Theta)^4 < (\nabla_q \circ )^2 > + \propto (k\Theta)^{4,5} < \circ\nabla_q \circ > \propto (k\Theta)^4$$

it follows that

$$\lim_{\Theta\to 0} < N_1^{\;2} >^{\frac{1}{2}} \cong < (\nabla_q \circ F^*)^2 >^{\frac{1}{2}} +2 < (\nabla_q \circ F^*)\text{¦}_{(q,t,\Theta)} >^{\frac{1}{2}} \text{¢}(p - \nabla_q S)^{\frac{1}{2}}$$

$$+ < \text{¦}_{(q,t,\Theta)}^{\;\;2} >^{\frac{1}{2}} \text{¢}(p - \nabla_q S) + \approx < \text{¦}_{(q,t,\Theta)}^{\;\;2} >^{\frac{1}{2}} \text{¢}(p - \nabla_q S)$$

since $< \text{¦}_{(q,t,\Theta)}^{\;\;2} > \propto (k\Theta)^2$, and finally that

$$\lim_{\Theta\to 0} N_1 \cong \text{¦}_{(q,t,\Theta)} \text{¢}(p - \nabla_q S) \qquad \text{(D.46)}$$

that supports the approximated expression (37) from which it follows that $F^*_1 = 0$.

# Appendix E

If we consider a system (e.g., an ideal gas) at equilibrium in a fluctuating vacuum environment in a container of side $\Delta L$ we have two way to calculate the system energy fluctuations. One is the standard one given by the statistical physics where the energy fluctuations are function of the temperature T, the second one is given by using the SQHA approach that will furnish an expression function of the amplitude $\Theta$ of the vacuum noise. Since the result must be same because independent by the way it is calculated, a connection between the temperature of a gas at equilibrium in a vacuum environment and the amplitude $\Theta$ of its fluctuations can be established.
To this end, let's start by calculating the energy fluctuations amplitude in the SQHA notations

$$< \underline{E}, \underline{E} >=< \int ... \int ° \, \mathrm{H}_{(q,p)} d^{3n} p, \int ... \int ° \, \mathrm{H}_{(q,p)} d^{3n} p > \tag{E.1}$$

By using the identity $°_{(q,p,t)} = \mathrm{n}_{(q,t)} ¢ ( p - \nabla_q S )$ (introducing (111) in (A.1) at zero order for small $\Theta$) we obtain

$$< \underline{E}, \underline{E} >=< \mathrm{n}_{(q,t)} \int ... \int \delta(p - p_{qu}) \mathrm{H}_{(q,p)} d^{3n} p, \mathrm{n}_{(q,t)} \int ... \int \delta(p - p_{qu}) \mathrm{H}_{(q,p)} d^{3n} p >$$

(E.2)

$$\begin{aligned} < \underline{E}, \underline{E} > &=< \mathrm{n}_{(q,t)} \mathrm{H}_{(q,p_{qu})}, \mathrm{n}_{(q,t)} \mathrm{H}_{(q,p_{qu})} > \\ &=< ¦_{(q,\Theta)} \mathrm{H}_{(p_{qu})}, ¦_{(q,\Theta)} \mathrm{H}_{(p_{qu})} > (t - t_0) \\ &= (3n) \underline{\mathrm{H}}^2 < ¦_{(q,\Theta)}, ¦_{(q,\Theta)} > (t - t_0) \end{aligned} \tag{E.3}$$

where the system volume has been assumed unitary and where

$$\mathrm{H}_{(q,p_{qu})} = \int ... \int \delta(p - p_{qu}) \mathrm{H}_{(q,p)} d^{3n} p = \sum_{\sim =1}^{3n} \frac{p_{qu\sim} \, p_{qu\sim}}{2m}. \tag{E.4}$$

is brought out the variance operation since is not function of the space being the quantum potential of a pure sine or cosine wave function constant.
Moreover, by introducing the notation

$$< ¦_{(q,\Theta)}, ¦_{(q,\Theta)} >=< ¦_0, ¦_0 > k\Theta \tag{E.5}$$

where by (61)

$$< ¦_0, ¦_0 > \propto \frac{1}{\mathrm{a}_c^{\;2}} \tag{E.6}$$

we obtain

$$\Delta E = V < \underline{E}, \underline{E} >^{1/2} = V (3n)^{1/2} \underline{\mathrm{H}} < ¦_0, ¦_0 >^{1/2} (¿ \, k\Theta)^{1/2} \tag{E.7}$$

where

$$\underline{\mathrm{H}} = \{ (3n)^{-1} \sum_{\sim =1}^{3n} \frac{p_{qu\sim} \, p_{qu\sim}}{2m} \}^{1/2} = \frac{< p^2_{\sim} >}{2m} \tag{E.8}$$

and where, conceptually, $\tau$ = $(t\text{-}t_0)$ is the interval of time along which the vacuum fluctuations are observed. Since for a particle of an ideal gas the time for a free path can be at maximum the interval of time between two collisions with the vessel walls (e.g., a cube of side ΔL) it follows that the maximum time τ of fluctuations observation reads

$$2 = \frac{m\,\Delta L}{< p^2_{\sim} >^{1/2}} = \frac{m\,\Delta L}{(2m\underline{H})^{1/2}} .$$

(E.9)

Moreover, by equating the energy fluctuations of *n* independent point mass particles of the ideal gas at equilibrium [65], given by the formula

$$\Delta E = (C_v)^{1/2} kT = (\frac{3n}{2})^{1/2} kT \qquad \text{(E.10)}$$

with (E.7), it follows that

$$\Delta E = \Delta L^3 (3n)^{1/2} \underline{H} < ¦_0 , ¦_0 >^{1/2} (2\; k\Theta)^{1/2} = (\frac{3n}{2})^{1/2} kT \qquad \text{(E.11)}$$

where $V$ is the volume of the system.
The value of H depends by the state of the system and for T close to the absolute zero it can be easily calculated.
Even if Θ does not coincide with the thermodynamic temperature T, going toward the absolute null temperature by steps of thermodynamic equilibrium, correspondingly, Θ must decrease to zero since the systems fluctuations must vanish both as a function of Θ as well as a function of T. Therefore, in this case we expect that

$$\lim_{T \to 0} \Theta = 0 \qquad \text{(E.12)}$$

We suppose the gas sufficiently rarefied so that N/V is very small and the Bose-Einstain condensation temperature $T_{BE}$ is smaller than that one of the probing ideal gas temperature.
The value of $\lim_{T \to 0} H_{(p)}$ depends by the state of the system. When the limit of Θ→0 is achieved at thermodynamic equilibrium and all the particles are in the fundamental state, at T=0 in a vessel of side ΔL, it follows that

$$\lim_{T \to 0} \underline{H} = \frac{®^2 \hbar^2}{8m\Delta L^2} , \qquad \text{(E.14)}$$

that

$$2 = \frac{2m\,\Delta L^2}{®\hbar} \qquad \text{(E.15)}$$

and that

$$\lim_{T \to 0} \Delta E = \Delta L^3\; (3n)^{1/2} \frac{®^2 \hbar^2}{8m\Delta L^2} \lim_{T \to 0} < ¦_0 , ¦_0 >^{1/2} (2\; k\Theta)^{1/2} = (\frac{3n}{2})^{1/2} kT \qquad \text{(E.16)}$$

.
Moreover, given that by (E.6), as can be checked at the end, it holds

$$< ¦_0 , ¦_0 > \propto \frac{1}{a_c^{\;2}} \propto k\Theta , \qquad \text{(E.17)}$$

the energy fluctuations are linear both as a function of Θ as well as of T . Therefore for an ideal gas (E.12) coherently reads

$$\lim_{T \to 0} \Theta \propto T = 0 \qquad \text{(E.18)}$$

and hence, we can redefine $\Theta$ by a proportionality constant to have

$$\lim_{T \to 0} \Theta = T \qquad \text{(E.19)}$$

Thence, if we measure Θ in a scale such as Θ = T , we obtain

$$(\frac{k\Theta}{2})^{1/2} = V \frac{®^{2}\hbar^{2}}{8m\Delta L^{2}} \lim_{T \to 0} < ¦_{0}, ¦_{0} >^{1/2} \, 2^{\; 1/2} \qquad \text{(E.20)}$$

that by re-writing (61) as

$$< ¦_{0}, ¦_{0} > = \frac{\underline{«}}{a_{c}^{\;2}} \qquad \text{(E.21)}$$

leads to

$$\lim_{\Theta \to 0} a_{c}^{\;2} = \frac{®^{3}\hbar^{3}\Delta L^{4}}{mk\Theta\, 2^{4}} \underline{«} \quad . \qquad \text{(E.22)}$$

Moreover, given that the term $\frac{\Delta L^{2}}{\hbar} = «$ has the dimension of a mobility, it follows that the term

$$\underline{«}\hbar\Delta L^{4} = \frac{«\hbar}{\Delta L^{2}} , \qquad \text{(E.23)}$$

(E.24)

has the dimension of a number and, hence, it follows that

$$a_{c} = (\underline{«}\hbar\Delta L^{4})(\frac{®}{2})^{3/2} \frac{\hbar}{(2mk\Theta)^{1/2}} . \qquad \text{(E.25)}$$

Finally, in order to define the numerical value of $(\underline{«}\hbar\Delta L^{4})$ we need an additional information. Since the length $a_{c}$ defines the maximum of quantum state delocalization, it is linked to the indetermination principle. It is matter of fact that smaller is the value of $a_{c}$ larger are is the energy of vacuum fluctuations (and hence the connected momentum variance).

Moreover, given that on distance shorter that $a_{c}$ for any system it holds the wave-particle equivalence and hence we cannot perturb a part of it without disturbing all the system (non local behavior), to perform a statistical measurement the system and the measuring apparatus must be far apart a distance larger that $a_{c}$. This fact influences both the time of measurement as well as its precision. Therefore the numerical value of $(\underline{«}\hbar\Delta L^{4})$ is determined by the experimental value of the physical uncertainty.

This can be ascertained as follow: given that the time for traveling the distance $a_{c}$ is $2_{c} = \frac{a_{c}}{c}$, for performing a statistical measurement between quantum uncorrelated systems we need a time $\Delta t > {}^{2}{}_{c}$. Moreover, given the relativistic energy of a particle of mass $m$ in presence of vacuum energy fluctuations $\Delta E_{(\Theta)}$ such as $< \Delta E_{(\Theta)} >= k\Theta$, for the classical case ($mc^2 >> k\Theta$) we obtain

$$\Delta E \approx (< (mc^2 + \Delta E_{(\Theta)})^2 - (mc^2)^2 >)^{1/2} \cong (< (mc^2)^2 + 2\Delta E_{(\Theta)} - (mc^2)^2 >)^{1/2}$$

$$\cong (2mc^2 < \Delta E_{(\Theta)} >)^{1/2} \cong (2mc^2 k\Theta)^{1/2}$$

from which, by imposing the uncertainty relation that reads

$$\Delta E \Delta t > \Delta E \Delta {}^{2}{}_{c} = \frac{(2mc^2 k\Theta)^{1/2} \text{a}_c}{c} = \frac{h}{2},$$

it follows that

$$( \text{«} \hbar \Delta L^4 ) = 2(\frac{®}{2})^{-1/2} .$$

For $\Theta$ that goes to zero (i.e., standard quantum mechanics) the measuring time goes to infinity and the energy fluctuation $\Delta E$ goes to zero (we have a perfect overall quantum system with exactly defined energy levels). This makes clear that in a perfect quantum universe the measuring process is endless (i.e., not possible) confirming that the classical behavior is needed to the definition of the quantum mechanics based on the measuring process.

The minimum uncertainty principle comes by the fact that below the length $\text{a}_c$ the locality is lost (I cannot divide such a system into parts in order to improve my precision) united to the fact that I cannot collect information (ultimately to make a measure) in a time shorter than that one needed to the interactions and information to travel such distance.

By using (in the limit of small $\Theta$) the convention that $\Theta$ equals the temperature of an ideal gas at equilibrium (E.19) and by imposing that the physical uncertainty principle is verified, $\text{a}_c$ remains defined by Formula (E.??). Moreover, by measuring the temperature of an ideal gas at equilibrium in a fluctuating vacuum environment, we detect the value of $\Theta$ of the theory.

# Appendix F

*Large-scale quantum potential*

If we write the WFM |ψ| in the form

$$\lim_{|q|\to\infty} |\cdot| = \lim_{|q|\to\infty} n^{1/2} = M_{(q)} exp[-f(q)] , \quad \text{(F.1)}$$

with the sufficiently general condition

$$\lim_{|q|\to\infty} f(q) > 0 \quad \text{(F.2)}$$

it follows that for a finite $\text{a}_L$ condition (??) that implies that

$$\lim_{|q|\to\infty} -\nabla_q V_{qu} = \lim_{|q|\to\infty} \nabla_q (\frac{\hbar^2}{2m}) n^{-1/2} \nabla_q \bullet \nabla_q n^{1/2} = 0 \quad \text{(F.3)}$$

leads to

$$\lim_{|q|\to\infty} \nabla_q M_{(q)}{}^{-1} exp[f(q)] \nabla_q \bullet \nabla_q M_{(q)} exp[-f(q)] = 0 . \quad \text{(F.4)}$$

The condition (above) implies that the following differential equalities must contemporarily be verified

$$\lim_{|q|\to\infty} \nabla_q(\nabla_q \bullet \nabla_q f(q)) = 0 \tag{F.5}$$

$$\lim_{|q|\to\infty} \nabla_q(\nabla_q f(q) \bullet \nabla_q f(q)) = 0 \tag{F.6}$$

$$\lim_{|q|\to\infty} \nabla_q M^{-1}(q)(\nabla_q \bullet \nabla_q M(q)) = 0 \tag{F.7}$$

$$\lim_{|q|\to\infty} \nabla_q M^{-1}(q)(\nabla_q f(q) \bullet \nabla_q M(q)) = 0 \tag{F.8}$$

The system of equations (F.5, F.8) has no simple and immediate solutions in three dimensional space. For sake of simplicity, here we give the solution for the mono-dimensional case, since for many practical cases (e.g., the interaction of a couple of particles, as in a real gas or in a chain of neighbors interacting atoms) this is still of great interest. In this case (F.5, F.8) reads

$$\lim_{|q|\to\infty} \frac{d}{dq}[(\frac{df(q)}{dq})^2] = 0, \tag{F.9}$$

$$\lim_{|q|\to\infty} \frac{d}{dq}[(\frac{d^2 f(q)}{dq^2})] = 0, \tag{F.10}$$

$$\lim_{|q|\to\infty} \frac{d}{dq}[M^{-1}(q)(\frac{df(q)}{dq})(\frac{dM(q)}{dq})] = 0, \tag{F.11}$$

$$\lim_{|q|\to\infty} \frac{d}{dq}[M^{-1}(q)(\frac{d^2 M(q)}{dq^2})] = 0. \tag{F.12}$$

If we approximate $f(q)$ for large |q| by a polynomial expression of maximum degree *h* such as

$$\lim_{|q|\to\infty} f(q) \approx P^h(q) \tag{F.13}$$

we obtain

$$h < \frac{3}{2} \tag{F.14}$$

If we are interested in bounded or localized states (owing the property $\lim_{|q|\to\infty} |\cdot| = 0$), that requires $\lim_{|q|\to\infty} |f(q)| = +\infty$ and hence that

$$\lim_{|q|\to\infty} |f(q)| \approx |P^h(q)| \propto |q^h| , \tag{F.15}$$

it necessary follows that

$$0 < h < \frac{3}{2}\,. \tag{F.16}$$

If we want to comprehend the case $\lim_{|q|\to\infty} |\cdot| = constant$, the equality must added to condition (F.16.a), to obtain

$$0 \le h < \frac{3}{2} \tag{F.17}$$

As far as it concerns (14.4 -5), if we approximates $M_{(q)}$ for large |q| by a polynomial expression such as

$$\lim_{|q|\to\infty} M_{(q)} \cong q^m \sum_n a_n\, exp[iA_n{}^p{}_{(q)}] \tag{F.18}$$

where $A_n{}^p{}_{(q)}$ is a polynomial expression of maximum order *p*, conditions (F.9, F.12) are contemporarily satisfied by $m \in \Re$ and $p \le 1$.
In the case of attractive long-range (smooth) forces, whose zero level can be posed to infinity for which it holds

$$\lim_{|q_\sim|\to\infty} A_n{}^p{}_{(q)} \propto q \tag{F.19}$$

(as for L-J type potentials where the proportionality constant reads $(2mE)^{½}$) we have $p = 1$ and, hence, (14.1) is warranted just by the condition

$$0 \le h < \frac{3}{2} \tag{F.20}$$

## Nomenclature